\documentclass[a4paper]{article}

\usepackage{amsfonts,amssymb,amsthm,epsfig}
\usepackage[totalwidth=17cm,totalheight=24cm]{geometry}

\newcommand{\C}{{\mathbb C}}

\newcommand{\cD}{{\mathcal D}}

\newcommand{\cS}{{\mathcal S}}

\newcommand{\be}{\begin{eqnarray}}
\newcommand{\ee}{\end{eqnarray}}

\newcommand{\su}{{\mathfrak su}}
\newcommand{\mb}{\bar{m}}
\newcommand{\tli}{\tilde{i}}
\newcommand{\tlo}{\tilde{o}}
\newcommand{\tlX}{\tilde{X}}

\begin{document}

\title{Non-Metric Gravity I: Field Equations}

\date{v3: Oct 15, 2007}

\author{Kirill Krasnov 
\thanks{School of Mathematical Sciences, University of Nottingham, Nottingham,
  NG7 2RD, UK and Perimeter Institute for Theoretical Physics, 
Waterloo, N2L 2Y5, Canada.}}

\maketitle

\begin{abstract} We describe and study a certain class of modified
gravity theories. Our starting point is Pleba\'nski formulation of
gravity in terms of a triple $B^i$ of 2-forms, a connection $A^i$ and a
``Lagrange multiplier'' field $\Psi^{ij}$. The generalisation we consider stems 
from presence in the action of an extra term proportional to a scalar
function of $\Psi^{ij}$. As in the usual Pleba\'nski general relativity (GR) case, 
a certain metric can be constructed from $B^i$. However, unlike in GR, the 
connection $A^i$ no longer coincides with the self-dual part of the
metric-compatible spin-connection. Field equations of the theory
are shown to be relations between derivatives of the metric and components
of field $\Psi^{ij}$, as well as its derivatives, the later being in contrast to the
GR case. The equations are of second order in derivatives.
An analog of the Bianchi identity is still present in the theory,
as well as its contracted version tantamount to energy
conservation equation. 
\end{abstract}

\maketitle

\section{Introduction}

Modified gravity theories have become popular recently, motivated mainly
by the fact that Einstein's general relativity (GR) interprets the available
observational data in terms of most of the content of the universe
being in the form of some dark energy and matter that have never been 
observed directly. Modified gravity theories are essentially of two main types.
One type introduces new fields. Some examples
of these are: Brans-Dicke \cite{Dicke} 
(a generalisation of GR in which the gravitational
constant is replaced by a scalar field), a scalar-vector-tensor theory of Bekenstein
\cite{Bekenstein:2004ne} that is designed to give modified Newtonian
dynamics (MOND) as its non-relativistic limit, the non-symmetric gravitation theory of
Moffat \cite{Moffat:1995fc} that is a theory with non-metric gravitational degrees of freedom,
a scalar-vector-tensor theory of Moffat, see e.g. \cite{Clayton:1999zs}, 
brane-world theories \cite{Dvali:2000hr}, \cite{Shtanov:2000vr} that introduce an extra 
spacetime dimension (an thus an infinite number of new fields). 
Another interesting recent proposal \cite{Punzi:2006nx} calls for
gravity to be described by the so-called area metric instead of the metric in the usual sense. 
One of the major challenges for all these models
is to show how Einstein's GR arises in ``usual circumstances''. The other
type of modifications considered is that of pure GR, i.e. with no
new fields added, but with modified field equations. The examples of such
modifications considered so far all involved higher derivatives. Indeed, it
is well-known that Einstein's GR is the only diffeomorphism-invariant theory in which
the main dynamical field is the spacetime metric and which has at most second order 
derivatives in the action. Therefore, in order
to modify GR without introducing new fields there is no other way but to 
introduce higher derivative terms in the action. These modifications
typically also increase the number of degrees of freedom described by the theory,
as one now needs to specify more initial data (more than just first time derivatives
of the spatial metric). The quantum gravity induces modifications (quantum corrections) 
are precisely of this type, which is one of the reasons for interest in this models.
Some examples of these are: theories used for string cosmology, 
see e.g. \cite{Biswas:2006bs}, Weyl conformal gravity 
considered in e.g. \cite{Mannheim:1992vj}, the so-called $f(R)$ theories
\cite{Carroll:2003wy}. There are other
modified gravity theories considered in the literature, and the
above list makes no attempt to be complete. The two types described are
not necessarily mutually exclusive, as integrating the new fields out
in a theory of the first type one typically gets a theory of the second type 
(typically with an infinite number of derivatives - a non-local theory).
One of the lessons learnt from all the above developments is that
it is rather hard to modify Einstein's general relativity in
any interesting way and to remain consistent with what is observed.

In this paper we introduce and study a new class of modified gravity theories
which in a certain sense falls simultaneously into both of the categories described above. 
Thus, to formulate the theory certain new fields are added to the gravity action.
However, this is done in such a way that the new fields do not propagate. Another,
more precise way to describe the class of theories considered in this paper is
as follows. General relativity can be described \cite{Plebanski} as a
certain topological field theory called BF theory (with
no propagating degrees of freedom) in which some of the large set of symmetries
that make this theory topological have been ``removed'' or ``gauge fixed'' by
adding to the action certain Lagrange multiplier terms. The theories we
consider are exactly of the same type, the difference with GR being that
the ``topological symmetries'' are removed in a certain modified fashion.
In other words, a more precise description of the theories we consider is not that 
new non-propagating degrees of freedom are added as compared to gravity, 
but that the topological symmetries that are to be removed to get a theory
with propagating degrees of freedom are ``gauge fixed'' in a modified fashion.
Thus, what this paper describes is the set of generally covariant modifications 
of four-dimensional GR that propagate exactly two degrees of freedom. Only very
few such generalizations were known in the past, see the end of this Introduction
for references. This work provides a construction of an infinite parameter
family of such theories. The above description makes it clear that the type of 
modification we consider is unlike what is done in most if not all previously 
considered modified gravity scenarios. 

Yet another description of the effect of the modification is as follows. 
As in Pleba\'nski formulation of GR \cite{Plebanski}, a certain new field, referred to as the Lagrange 
multiplier field $\Psi$ is introduced. In the GR case
some of the equations of motion of the theory express $\Psi$ 
in terms of second derivatives of the metric (or, in terms
of the curvature of $g$). In modified theories we consider $\Psi$
is no longer so simply expressible through $g$. The equation
that relates $g$ and $\Psi$ is, schematically
\be\label{intr-1}
D^2g + \tilde{D}^2 f(\Psi) = \Psi + \phi(\Psi),
\ee
where $D^2,\tilde{D}^2$ are certain second-order differential
operators, and $\phi,f$ are certain functions of $\Psi$.
This equation can be solved, at least in principle, for
$\Psi$, with the solution being, again schematically
\be
\Psi = D^2 g + \sum_{n,m=1}^\infty a_{nm} (\tilde{D}^2)^n (D^2 g)^m,
\ee
where $a_{nm}$ are some coefficients. Substituted back into
the action this gives the usual Einstein-Hilbert functional 
corrected by an infinite series of higher order terms in the curvature
of $g$.

At the very least, the class of models we consider is an interesting
compact (and useful for quantum considerations, see more
comments on this below) way of re-writing such infinite expansions.
As is clear from (\ref{intr-1}) the coupled equations for $g,\Psi$
are second order, which guarantees that many 
unpleasant problems with higher derivative theories are automatically avoided. Moreover,
as we have already mentioned above, the new fields $\Psi$ are added in such a way
that the structure of the phase space of the theory is unchanged. Thus, there
are no new degrees of freedom as compared to the usual GR case, even though this is
not at all obvious by examining (\ref{intr-1}), as it might seem that one has
to specify the initial data for both $g$ and $\Psi$. As a detailed analysis shows,
see the recent work of Bengtsson \cite{Bengtsson:2007zx}, and as we have already
commented on above, this is not the case: the count of the degrees of freedom is
unchanged as compared to GR. Thus, the class of theories we consider
is quite interesting in that it does not increase the number of derivatives in the
field equations, as well as does not introduce any new degrees of freedom apart
from those already present in GR. The price one has to pay for this is that the
theory becomes non-metric.

\bigskip
\noindent{\bf ``Renormalizable'' gravity.}
Most optimistically, this formulation of (quantum corrected) gravity could allow
one to come to terms with the behaviour of gravity under renormalization, as we shall
now review. It was argued in \cite{Krasnov:2006du} that in the Pleba\'nski 
formulation \cite{Plebanski} the behaviour of gravity under
renormalization is more transparent and simple  than in 
the usual metric based scheme. In particular, it was argued that
the renormalization group flow is a flow in the space of
scalar functions of two complex variables. Thus, using certain 
field redefinition arguments as well as direct computation (based on a certain
gauge fixing procedure) it was argued that the class of theories defined by the action:
\be\label{action}
S[B,A,\Psi]= \frac{1}{8\pi G} \int_M B^i \wedge F^i - \frac{1}{2} 
\left( \Lambda \delta^{ij} + \Psi^{ij} + 
\phi \delta^{ij} \right) B^i \wedge B^j, \\ \nonumber
\phi=\phi({\rm Tr}(\Psi^2),{\rm Tr}(\Psi^3)) 
\ee
is closed under the renormalization group flow. It is in this sense
that the theory is referred to as ``renormalizable''. As explained
above, for any $\phi$ the theory (\ref{action}) is actually 
equivalent to a metric theory given by an infinite expansion in powers
of the curvature of this metric. Thus, the statement that
(\ref{action}) is closed under the renormalization group flow does not
contradict what is known from the usual perturbative quantum gravity.
In this action $B^i$ is a (complex) $\su(2)$ Lie-algebra valued 
2-form (indices $i,j,\ldots=1,2,3$ are
$\su(2)$ Lie-algebra ones), $F^i=dA^i + (1/2)[A,A]^i$ is the curvature of
an $\su(2)$ Lie-algebra valued connection $A^i$, $\Psi^{ij}$ is a 
traceless symmetric ``Lagrange multiplier'' field, that on shell
gets related to the ``curvature'', see more on this below. Let us note that
instead of requiring $\Psi^{ij}$ to be traceless one could add to the action
a Lagrange multiplier enforcing this condition. Continuing with our
description of the above action, $\Lambda$ is a multiple of the cosmological constant, 
the ``usual'' cosmological constant $\tilde{\Lambda}$ 
appearing in the EH action in the combination $(R-2\tilde{\Lambda})$ is related
to the one in (\ref{action}) via $\tilde{\Lambda}=3\Lambda$, and $G$ is
the Newton's constant. Finally, the function 
$\phi({\rm Tr}(\Psi^2),{\rm Tr}(\Psi^3))$ is (as its arguments indicate) 
a function of two scalars that can be constructed from $\Psi^{ij}$. The fact
that this function can only depend on two scalar invariants, not three as it
would be for a general $3\times 3$ matrix, is a consequence of the tracelessness of $\Psi$.
The function $\phi$ is zero for Pleba\'nski formulation of GR \cite{Plebanski}. 
Equivalently, one can view the usual cosmological constant as a function
$\phi(\Psi)$ that is independent of $\Psi$ (constant).
As is shown in \cite{Krasnov:2006du}, the additional $\phi$-term is generated
by quantum corrections, and, importantly, it is argued that this is the only term that gets generated.
The renormalization group flow is then a flow in the space of functions $\phi$, which 
makes this function scale dependent: $\phi=\phi_\mu$, where $\mu$ is energy
scale. The asymptotic safety scenario of Weinberg \cite{Weinberg} 
can be reformulated in this context as a conjecture that there exists a 
non-trivial limit $\phi^*=\lim_{\mu\to\infty} \phi_\mu$, 
and that the theory that describes gravity at our energy scales 
is on a renormalization group flow trajectory that leads to $\phi^*$. 
However, gravity being a diffeomorphism invariant theory, and there being
no way to define what ``energy scale'' means in a diffeomorphism invariant
context (at least in the absence of other fields that can provide one with a reference background), 
it is not unreasonable to suppose that the theory that describes
gravity at {\it all} scales is the one with $\phi=\phi^*$, i.e. 
that the theory at all scales coincides with the UV fixed point one.
The main aim of this paper is to study the theory (\ref{action})
assuming that $\phi$ is a given function (equal to the unknown $\phi^*$). 
The asymptotic safety conjecture and the problem of determining 
$\phi^*$ is not addressed here. 

\bigskip
\noindent{\bf Quantum corrections or a new scale?} Before we embark on
a systematic study of the theory defined by (\ref{action}) it is
worth emphasising one important conceptual difference between 
(\ref{action}) and the usual ``quantum corrected'' Einstein-Hilbert
action. Thus, let us remind the reader that Einstein-Hilbert action
receives quantum corrections from counterterms necessary to
cancel the divergences arising in perturbation theory. This
counterterms have the form of various invariants constructed from 
Riemann curvature tensor integrated over the spacetime. At leading
(first) order the quantum corrections are of the form $(curvature)^2$ integrated
over the spacetime. As curvature has dimensions $1/L$, $L$ being length,
such terms are dimensionless. To 
give them the dimension $M\cdot L$, $M$ being mass,
required from the action one has to multiply these terms by a dimensionfull
parameter - the Planck constant $\hbar$. Thus, the terms in the quantum corrected action that are
of second order in curvature have a multiple of $\hbar$ in front
of them and thus are {\it quantum} corrections. This power of $\hbar$ agrees
with the fact that these terms arise at one loop order of perturbation theory.
Containing a prefactor of $\hbar$, these terms should be
ignored when considering the classical gravity theory, as this is obtained
via $\hbar\to 0$ limit. This nicely complements the fact that, had we
considered such terms in the classical action, the resulting theory
would have fourth derivatives of the dynamical fields, and as such
would have various unpleasant problems. This provides us with an
``explanation'' of why Einstein-Hilbert action is the correct
action for the classical theory of gravity.\footnote{There is another,
more standard, and actually related explanation of this that
uses the Wilsonian renormalization group flow arguments, see e.g.
\cite{Weinberg}. We will not use such Wilsonian way of reasoning in the
present ``classical'' paper.}

Let us now consider the theory defined by (\ref{action}). In spite of
the seeming similarity to the usual metric based quantum gravity 
in that all powers of the ``curvature field' $\Psi$ arise 
in counterterms (in (\ref{action}) the
function $\phi$ may be thought of as a powers series expansion in both of
its arguments), the theory (\ref{action}) is quite different. Indeed, in 
the usual metric based quantum gravity higher powers of the curvature
field imply higher derivatives in the equations of motion. In the case of
(\ref{action}) the field $\Psi$ becomes related to the curvature only
on-shell, and the action is second-order for any choice of the function 
$\phi$. This suggests that the $\phi$-term of the action (\ref{action})
should be considered not a quantum correction, but instead an additional term
in the classical action. Note, also that the two expansions: one in powers of the
curvature in the usual metric based gravity, and the other in powers of $\Psi$ in 
theory (\ref{action}) are quite different. Indeed, as is clear from (\ref{intr-1}) 
even a seemingly innocuous modification with $\phi=q{\rm Tr}(\Psi)^2$ leads to an infinite 
expansion in powers of the curvature when interpreted in metric terms!

Let us supplement this ``second
derivatives only'' argument with dimensional analysis. The function
$\phi({\rm Tr}(\Psi^2),{\rm Tr}(\Psi^3))$ has the dimension of the other
terms in the brackets in the second term of (\ref{action}), i.e. $1/L^2$. 
This term becomes equally important as the term $\Psi^{ij}$ when 
the curvature field $\Psi^{ij}$ is of the same order as $\phi$. This can
be phrased in a more meaningful way by saying that the effects of the
$\phi$-term become significant for curvatures $\Psi^{ij}$ such that
the eigenvalues of the matrix $\Psi^{ij}$ are of the same order as the
function $\phi$ of these eigenvalues: 
$\lambda_{1,2} \sim \phi(\lambda_1,\lambda_2)$. This equation defines a
new length scale $l_*: \lambda_{1,2} \sim 1/l^2_*$. 

For ``small'' curvatures the term proportional to ${\rm Tr}(\Psi)^2$ in
the expansion of $\phi$ is dominant. The coefficient in front of this term
has dimensions $L^2$. It is this coefficient that can be used as a definition
of the length scale $l_*$:
\be
\phi({\rm Tr}(\Psi^2),{\rm Tr}(\Psi^3)) = \pm l_*^2 {\rm Tr}(\Psi)^2 + O(\Psi^3),
\ee
where, in principle, both signs are possible in the above formula.

Now, to determine whether the $\phi$-term in the action (\ref{action})
is a classically important term or just a quantum correction one must
ask whether $l_*$ goes to zero when $\hbar\to 0$. In the usual
metric based scheme $l_*\sim l_p$, the Planck length and does go to zero 
in the classical limit. In the pure gravity case, i.e. when no matter degrees of
freedom is present, there is no other scale but $l_p$, so $l_*$ can only be
a multiple of $l_p$. In this case there is no other choice but to interpret 
the theory (\ref{action}) as a quantum corrected gravity theory,
with the length scale $l_*$ being the Planck scale. The theory then describes 
how the familiar notions of geometry get modified when one approaches the Planck scale.
This interpretation has to be taken with caution, however, as there can only
be a limited applicability of a classical description in the essentially quantum regime
close to the Planckian scale. Now, when matter degrees of freedom are present
in the theory they do affect the running of $\phi$ and will generally introduce other scales. 
It is then possible that the modification of the
type we consider survives the classical limit. Indeed, the fact that (\ref{action}) is 
second-order in derivatives suggests that $l_*$ may remain finite even when
$\hbar$ is sent to zero. In other papers of the series we will follow up
on this idea and see what kind of modifications arise when the scale $l_*$ is
taken to be large (astrophysical). 

\bigskip
\noindent{\bf ``Non-metric'' modified gravity theory.}
Thus, if one is to take the quantum gravity scenario reviewed above 
seriously, one is led to consider the modified gravity theory (\ref{action}) 
(with some yet unknown function $\phi({\rm Tr}(\Psi^2),{\rm Tr}(\Psi^3))$ depending
on some length scales defined by the matter) as the classical limit $\hbar\to 0$ of the 
quantum corrected theory of gravity. The major difference between 
(\ref{action}) as a classical theory of gravitation and GR is that
the theory (\ref{action}) is no longer about metric in any
obvious way. Indeed, as in
Pleba\'nski formulation of GR, the gravity theory (\ref{action}) is
formulated in such a way that the metric never appears. However,
in Pleba\'nski gravity one of the equation implies ``metricity'',
which then in particular implies that $A$ in (\ref{action})
coincides with the metric-compatible connection for some metric $g$. 
In contrast, when $\phi\not=0$ it is no longer true
that $A$ is metric-compatible. 
It is for this reason that we propose to refer to this theory 
(or, rather to the class of theories defined by (\ref{action}))
as {\it non-metric gravity}. 

In order to avoid a possible confusion we would like to stress that the 
term ``non-metric'' is used here in the sense that no spacetime metric
appears in the formulation of the theory, not in the sense that some
additional ``non-metric gravitational'' degrees of freedom are present
as in e.g. \cite{Moffat:1995fc}.

As it should have become clear from arguments
above, Einstein's GR is a very good approximation to theories
(\ref{action}) when curvatures are smaller than $1/l_*^2$. 
However, when curvatures are large so are deviations from general 
relativity. In particular, deviations from Einstein's GR are
expected to be large near spacetime singularities. As we shall see
in one of the subsequent papers, the behaviour of all the fields of
theory (\ref{action}) near e.g. a would be singularity inside a black hole is
much less dramatic than that in Einstein's GR. This gives yet
another motivation to take the theories (\ref{action}) seriously. 

\bigskip
\noindent{\bf More on the nature of modification.}
It is worth explaining the main results of this paper in simple
terms, so that the reader is not lost in the rather technical
discussion of the main body of the paper. In order to 
explain what is the main feature that makes gravity (\ref{action}) different
from the usual GR let us recall some basic facts about the 
self-dual formulation of gravity due to Pleba\'nski \cite{Plebanski}.
In this approach, the main field that replaces the usual
metric of GR is a 2-form $B^i$ with values in the Lie algebra
of (complexified) $\su(2)$. One of the equations of the theory
states:
\be\label{metr-usual}
B^i\wedge B^j = \frac{1}{3} \delta^{ij} B^k\wedge B_k,
\ee
or, when written in the spinor form
\be
B^{AB}\wedge B^{CD} = \frac{1}{6} 
\epsilon^{A(C} \epsilon^{|B|D)} B^{EF}\wedge B_{EF}.
\ee
It is not hard to show that this equation implies that there exists a
quadruple of one-forms $\theta^{AA'}$ ($A,A'$ are spinor indices, 
see the main text for more details) such that $B^{AB}$ is given by:
\be\label{b-simple}
B^{AB} = \frac{1}{2} \theta^{AA'}\wedge \theta^B_{A'}.
\ee
In turn, the tetrad $\theta^{AA'}$ allows one to construct a
metric $ds^2 = \theta^{AA'}\otimes\theta_{AA'}$. 
It is in this sense that the equation
(\ref{metr-usual}) implies ``metricity'' of $B$. One can also
show that $B^{AB}$ given by (\ref{b-simple}) is self-dual
as a 2-form, where the self-duality is defined with respect to the above metric. 
Thus, after the ``metricity'' equation (\ref{metr-usual}) of 
Pleba\'nski formulation of GR is solved,
the field $B^{AB}$ becomes a self-dual 2-form, valued in $\su(2)$.
As such it can be thought of as a map identifying the space
of self-dual 2-forms (at a point of spacetime) 
with the Lie algebra $\su(2)$. The spinor formalism, which is
used to convert spacetime indices into spinor ones,
then interprets $B^{AB}$ as the identity map.

In gravity theory (\ref{action}) most of the above is still true.
The triple $B^i$ of 2-forms spans a 3-dimensional subspace in
the space of 2-forms, and this subspace can be called the space
of self-dual 2-forms. Demanding that there is a metric such
that this subspace is the space of self-dual 2-forms with respect
to this metric defines the metric modulo conformal factor, see
below for more on the ambiguities associated to this procedure
of determining the metric via $B^i$. Thus, $B^i$ becomes a map
from $\su(2)$ to the space of self-dual 2-forms of some metric. However, this map is no
longer the identity, see formula (\ref{b-spinor}) for the
corresponding expression. There are now at least two different, non-coinciding
ways of normalizing this map, which is related to the ambiguity in the choice
of the conformal factor for the metric. The action of the metric-compatible 
derivative operator $\nabla$ on 2-forms and the $A$-compatible
derivative operator $\cD$ on $\su(2)$-valued quantities is different,
see formulae (\ref{a-pm}), (\ref{a}) for the relations between
two derivative operators. This difference is quantified by departure of $B^{AB}$
from the identity map, which is in turn related to certain derivatives
of the field $\phi$, see relations (\ref{metricity-rels}).

The content of field equations of the theory is also similar to that in GR. 
The equations are given by (\ref{eqs-1}), (\ref{eqs-2})
and, similarly to the usual GR case, state that the anti-self-dual
part of the curvature 2-form $F(A)$ is proportional to the 
(traceless part of the) ``stress-energy'' 2-form $T$, while the self-dual part of 
$F(A)$ is related to the Lagrange multiplier field $\Psi$ as well as 
to the trace part of $T$ in a certain way. What is different is that one 
cannot anymore simply solve for the components of $\Psi$ in terms of the second derivatives
of the tetrad, as the corresponding equations become 
more involved and contain second derivatives of $\Psi$. 
One now has to solve the combined system of
equations for both the tetrad and the components of $\Psi$
simultaneously.

\bigskip
\noindent{\bf A purely metric formulation?}
Given the fact that a certain metric (defined by the tetrad)
does appear in the theory one may question if its legitimate
to refer to the theory as ``non-metric''. Indeed, as we have sketched above,
one might imagine solving for the components of the field $\Psi$ in terms of the
derivatives of the tetrad (one will generate an
infinite expansion in derivatives this way, as a derivative operator
will have to be inverted), and then substituting
the solution back into the action (\ref{action}). One would
obtain a generally covariant action for the tetrad-defined metric,
which would have the form of an infinite expansion in terms
of curvature invariants, with the first term being the usual
Einstein-Hilbert action. This is indeed possible in principle,
even though rather hard in practise. It would be
of interest to see which exactly subclass of such 
infinite expansions gets produced by actions of the form (\ref{action})
when $\phi$ is varied. This procedure seems to suggest
that the formulation (\ref{action}) is just a rather compact
way of re-writing a certain class of non-local 
(containing all powers of the curvature) gravity actions. The
critic may argue that this may be
interesting by itself, but probably does not call for a name
``non-metric'' gravity, which suggests a much more profound
change of conceptual framework. What we think makes 
this purely metric interpretation misleading are two things:
(i) the metric (or the tetrad) that is obtained in the 
process of solving the equations for $B$ is defined only up
to a conformal factor. We shall return to this point in the main text.
(ii) It is both possible and natural to couple matter fields not to 
a metric (which only arises when one of the equations of
motion is solved) but directly to the $B$ field. For example,
the action describing a coupling of Maxwell field to 
non-metric $B$ is given by \cite{CDJ}:
\be\label{em}
S_{EM}[B,a,\phi] =  \int_M f(a)\wedge \phi^i B^i - 
\frac{1}{2} \phi^i\phi^j B^i\wedge B^j,
\ee
where $a$ is the electromagnetic potential, $f(a)=da$, and
$\phi^i$ is a new 0-form field that is introduced to
make the coupling directly to $B$ possible, see the above
cited paper for more details on this action. Note that this action 
was introduced in \cite{CDJ} to describe the coupling of Maxwell
field to the usual metric gravity in the 2-form formalism.
However, this action does make sense even in the non-metric case,
as we shall demonstrate in another paper from the series. One could,
of course, couple the electromagnetic field directly to
the metric arising from $B$ in the usual way, but this
seems highly unnatural given the possibility of direct
coupling (\ref{em}). Moreover, such a direct coupling to the metric defined
using $B$ would make unavailable the possibility of shifts in $B$ that was heavily used in 
\cite{Krasnov:2006du} to describe renormalization. On the contrary, the action
(\ref{em}) being quadratic in $B$ allows for such shifts. This argument suggest
a strong criterion for possible matter couplings: the matter should only couple to the
$B$ field, and at most quadratic in $B$ couplings are allowed. Work is
currently in progress to find a description of the fermionic matter interacting with gravity 
that couples fermions directly to the $B$ field.

\bigskip
\noindent{\bf Energy conservation.} The theories we
consider, like GR, have Bianchi identity that can
be used to get energy conservation equations. In the retrospect,
this is not surprising, as energy conservation follows from
general covariance of the theory. However, given the
rather complicated nature of the generalisation, it
is somewhat miraculous that the energy conservation 
takes the form familiar from GR, as we shall see below.

\bigskip
\noindent{\bf Relation to ``neighbours of GR''.} An
interesting class of modified gravity theories motivated by the
pure connection formulation of gravity \cite{Capovilla:1991kx}
has been reviewed in e.g. \cite{Peldan:1993hi}. As described in 
\cite{Bengtsson:2007zx}, the class of theories we consider 
in the present paper is closely related to the theories in \cite{Peldan:1993hi}, 
our theories giving the most general modification of GR along these lines.

\bigskip
\noindent{\bf Organisation of the paper.} We start in the
next section by discussing the content of the modified ``metricity''
equations. Then, in section \ref{sec:spinors}, 
we develop spinor techniques for dealing with spacetime forms,
which will be of immense value for us in the following
sections. Section \ref{sec:comp} solves the equations
for the gauge field $A$, with the result being expressed
in terms of the usual metric-compatible spin-connection as well 
as the derivatives of the functions controlling non-metricity.
We discuss field equations in section \ref{sec:eqs}. Finally,
section \ref{sec:bianchi} derives and analyses the
``Bianchi'' identity. Appendices discuss some technical details
on ``metricity'' and ``compatibility'' equations. 

\section{Modified ``metricity'' equations}
\label{sec:metricity}

Our first goal is to study what the presence of 
the $\phi$-term in (\ref{action}) implies for the field $B^i$. We
remind the reader that in the absence of this term the equation
one obtains by varying (\ref{action}) with respect to $\Psi^{ij}$
guarantees ``metricity'', see (\ref{b-simple}). When $\phi$-term is present, 
this ``metricity'' equation gets modified, and it is our goal in this section to study what
this modification implies.

\bigskip
\noindent{\bf The ``metricity'' condition.}
Thus, let us consider the equation one gets by varying (\ref{action})
with respect to $\Psi^{ij}$. Special care should be taken in view of
the fact that $\Psi^{ij}$ is traceless. The equation one gets is as follows:
\be\label{metricity}
B^i\wedge B^j + \left( 2\partial_1 \phi \Psi^{ij} + 
3\partial_2 \phi ((\Psi^2)^{ij} - \frac{1}{3} \delta^{ij}{\rm Tr}(\Psi)^2)
\right) B^k\wedge B_k = \frac{1}{3} \delta^{ij} B^k\wedge B_k,
\ee
where $\partial_{1,2}\phi$ are the partial derivatives of $\phi$
with respect to the first and second arguments correspondingly. 
When $\phi=0$ we get the ``usual'' metricity equations of Pleba\'nski
formulation of GR. Before we can continue with our quest on 
what this equation implies we need to review some facts about
the field $\Psi^{ij}$. The simplest way to describe this field,
and relations satisfied by it is via the so-called spinor techniques,
which we therefore have to review.

\bigskip
\noindent{\bf Spinors.} We will not attempt a comprehensive introduction
to spinors, describing only what is relevant for us here. There are
many excellent sources, see e.g. \cite{Penrose}, on the subject, to which we 
refer the reader for more details. 

The action (\ref{action}), as well as the metricity equation
(\ref{metricity}) has been written using $\su(2)$ Lie-algebra 
valued fields. One can pass to the spinorial language by considering
the fundamental representation of this Lie-algebra in terms of
traceless $2\times 2$ matrices. Thus, each $\su(2)$ index $i,j,\ldots$ gets
replaced by a pair of spinor indices $AB$ where $A,B,\ldots = 1,2$. 
These indices are raised-lowered using the tensor $\epsilon^{AB}$, 
which is skew, i.e. $\epsilon^{BA}=-\epsilon^{AB}$. We will use 
conventions of \cite{Penrose}:
\be\label{raise-lower}
\psi^A = \epsilon^{AB} \psi_B, \qquad \psi_B = \psi^A \epsilon_{AB},
\ee
i.e. the index that is contracted is always located on top in the
first spinor and at the bottom in the second spinor. Due to the fact
that $\epsilon^{AB}$ is skew, the condition that a second rank spinor $X^{AB}$ 
is traceless is equivalent to the condition that it is symmetric. Thus, 
each $\su(2)$ index $i,j,\ldots$ is equivalent to a symmetric pair of
spinor indices $AB$. Note that the fact that $\epsilon^{AB}$ is skew
also implies that the product of any spinor with itself is zero:
$\lambda^A \lambda_A=0$.

\bigskip
\noindent{\bf Spinorial representation of the curvature field $\Psi$.}
The field $\Psi^{ij}$ is represented in the spinor form by a
rank 4 spinor $\Psi^{ABCD}$. This spinor is symmetric in each 
of the two pairs of indices $AB$ and $CD$ and is symmetric 
under the exchange of these two pairs (as $\Psi^{ij}$ is a 
symmetric matrix). It is easy to see that the condition that
$\Psi^{ij}$ is traceless is equivalent to the condition that
$\Psi^{ABCD}$ is completely symmetric. 

It is often convenient to choose a basis in the space of spinors. 
Let the basic spinors be denoted by $\tli^A, \tlo^A$. The reason why
we put a tilde over the basic spinors will become clear in the next
section, where we contrast the ``internal'' spinors that we are
dealing with in this section with ``spacetime'' spinors. The untilded
notation is reserved for the ``spacetime'' spinor basis.

It is customary
to normalize the basic spinors as:
\be
\tli^A \tlo_A = 1.
\ee
The curvature field $\Psi$ can then be decomposed as follows.
\be
\Psi^{ABCD} = \Psi_0 \tli^A \tli^B \tli^C \tli^D + 
\Psi_1 \tli^{(A} \tli^B \tli^C \tlo^{D)}
+ \Psi_2 \tli^{(A} \tli^B \tlo^C \tlo^{D)}+ 
\Psi_3 \tli^{(A} \tlo^B \tlo^C \tlo^{D)} +
\Psi_4 \tlo^A \tlo^B \tlo^C \tlo^D,
\ee
where the five quantities $\Psi_0,\ldots,\Psi_4$ are the (basis dependent)
spinor curvature components.

\bigskip
\noindent{\bf Basis of rank 2 spinors.} It is convenient to
introduce the following 3 basic rank 2 spinors:
\be\label{tl-X}
\tli^A \tli^B = \tlX_-^{AB}, \qquad \tlo^A \tlo^B = \tlX_+^{AB}, 
\qquad \tli^{(A} \tlo^{B)} = \tlX^{AB}.
\ee
The following commutational relations are easy to obtain:
\be
[\tlX_-,\tlX_+]=2\tlX, \qquad [\tlX,\tlX_+]=\tlX_+, \qquad [\tlX,\tlX_-]=-\tlX_-,
\ee
where our convention for the commutator is 
$[X,Y]_A^B = X_A^C Y_C^B - Y_A^C X_C^B$. This allows us to identify:
\be
\tlX_- = \left( \begin{array}{cc} 0 & 0 \\ -1 & 0 \end{array} \right),
\qquad 
\tlX_+ = \left( \begin{array}{cc} 0 & 1 \\ 0 & 0 \end{array} \right),
\qquad 
\tlX = \frac{1}{2} \left( \begin{array}{cc} 1 & 0 \\ 0 & -1 \end{array} \right).
\ee
The rank 2 spinors $\tlX_{\pm},\tlX$ 
thus span the $\su(2)$ Lie algebra. The Killing-Cartan form evaluated on
the basic rank 2 spinors is:
\be\label{x-product}
(\tlX_+,\tlX_-)=1, \qquad (\tlX,\tlX)=-\frac{1}{2},
\ee
where $(X,Y):=-{\rm Tr}(XY) = -X_A^B Y_B^A = X^{AB}Y_{AB}$.

\bigskip
\noindent{\bf Field $\Psi$ in terms of the rank 2 spinors.} The
following representation of the field $\Psi^{ABCD}$ in terms
of the rank 2 spinors $\tlX_{\pm}, \tlX$ turns out to be convenient:
\be
\Psi = \Psi_0 \tlX_-\otimes \tlX_- + 
\frac{\Psi_1}{2} (\tlX_-\otimes \tlX + \tlX\otimes \tlX_-)
+ \frac{\Psi_2}{6} ( \tlX_+ \otimes \tlX_- + \tlX_-\otimes \tlX_+ 
+ 4 \tlX\otimes \tlX) \\
\nonumber
+ \frac{\Psi_3}{2} ( \tlX_+\otimes \tlX + \tlX\otimes \tlX_+) 
+ \Psi_4 \tlX_+\otimes \tlX_+.
\ee
A proof is by an elementary computation.

\bigskip
\noindent{\bf Principal spinors and Petrov classification.}
Any completely symmetric rank 4 spinor $\Psi^{ABCD}$ can be represented as:
\be
\Psi^{ABCD} = k_1^{(A} k_2^B k_3^C k_4^{D)},
\ee
where $k_1^A,\ldots, k_4^A$ are referred to as {\it principal spinors}. The
{\it Petrov type} of the curvature field $\Psi^{ABCD}$ is determined
according to coincidence relations between the principal spinors. This goes
from type I - general, all 4 principal spinors are distinct to type N -
all 4 principal spinors coincide.

\bigskip
\noindent{\bf A convenient gauge.}
An elementary computation shows that by an ${\rm SU}(2)$ rotation of the 
basis $\tilde{i}^A, \tilde{o}^A$ the curvature field 
$\Psi^{ABCD}$ can be brought into the form:
\be\label{psi-gauge}
\Psi = \alpha (\tlX_-\otimes \tlX_- + \tlX_+\otimes \tlX_+ ) + 
\beta ( \tlX_+ \otimes \tlX_- + \tlX_-\otimes \tlX_+ + 4 \tlX\otimes \tlX),
\ee
i.e. the spinor curvature components $\Psi_1,\Psi_3$ can be eliminated
and the curvature components $\Psi_0,\Psi_4$ can be made equal. This
is possible always except in the case when $\Psi$ is of type $III$, i.e. when 
3 of the principal spinors coincide and do not coincide with the fourth 
principal spinor. A local ${\rm SU}(2)$ transformation allows to bring
the field $\Psi$ into the form (\ref{psi-gauge}) in any region
of spacetime where the Petrov type of $\Psi$ does not change. Let us
start by discussing the general case. The other (algebraically special) cases
(except type III) can be obtained from the general case sending
one of the parameters to zero, or making them equal. 
The type III case will be treated below separately.

\bigskip
\noindent{\bf Matrix products.} Using the gauge (\ref{psi-gauge}) it is
easy to evaluate the quantities ${\rm Tr}(\Psi)^2,{\rm Tr}(\Psi)^3$ that
we need as arguments of the function $\phi$ in terms of the quantities
$\alpha,\beta$. We have:
\be
{\rm Tr}(\Psi)^2 = 6\beta^2 + 2\alpha^2, \qquad 
{\rm Tr}(\Psi)^3 = -6\beta^3 + 6\beta\alpha^2.
\ee
Note that these formulae imply that $\phi$ is a function of $\alpha^2$ only.
Therefore, $\phi_\alpha|_{\alpha=0} = 0$, where $\phi_\alpha$ is a
partial derivative with respect to $\alpha$. This fact will be important
when we consider (in another paper) the spherically symmetric solution of the theory.

We will also need the traceless part of the matrix $\Psi^2$, as it
appears in the metricity equation (\ref{metricity}). We have:
\be
\Psi^2 - \frac{1}{3} {\rm Id} \, {\rm Tr}(\Psi)^2 = 2\alpha\beta 
(\tlX_-\otimes \tlX_- + \tlX_+\otimes \tlX_+ ) + \frac{1}{3}(\alpha^2-3\beta^2) 
( \tlX_+ \otimes \tlX_- + \tlX_-\otimes \tlX_+ + 4 \tlX\otimes \tlX).
\ee
Here ${\rm Id}$ is the identity tensor, which has the expression
\be\label{id}
{\rm Id} = \tlX_+ \otimes \tlX_- + \tlX_-\otimes \tlX_+ - 2 \tlX\otimes \tlX.
\ee
The tensor that appears in brackets on the left-hand-side of
(\ref{metricity}) is then given by:
\be\label{phi-1}
(2\alpha\partial_1\phi +
6\alpha\beta \partial_2\phi) (\tlX_-\otimes \tlX_- + \tlX_+\otimes \tlX_+ ) +
(2\beta\partial_1\phi + (\alpha^2-3\beta^2)\partial_2\phi) 
( \tlX_+ \otimes \tlX_- + \tlX_-\otimes \tlX_+ + 4 \tlX\otimes \tlX).
\ee

\bigskip
\noindent{\bf Change of coordinates.} It is convenient to use
$\alpha,\beta$ instead of the quantities ${\rm Tr}(\Psi)^2,{\rm Tr}(\Psi)^3$
as arguments of the function $\phi$. The change of coordinates is
elementary. Thus, we replace 
\be
\partial_1\phi = \partial_\alpha\phi \partial_1\alpha + 
\partial_\beta\phi \partial_1\beta, \qquad
 \partial_2\phi = \partial_\alpha\phi \partial_2\alpha + 
\partial_\beta\phi \partial_2\beta,
\ee
The following identities are then checked by an elementary
computation:
\be
2\alpha\partial_1 \alpha  + 6\alpha\beta\partial_2\alpha =1/2,  \qquad
2\beta \partial_1 \beta + (\alpha^2 - 3\beta^2) \partial_2 \beta = 1/6
\\ \nonumber
2\alpha\partial_1 \beta  + 6\alpha\beta\partial_2\beta = 0,
\qquad
2\beta \partial_1 \alpha (\alpha^2 - 3\beta^2) \partial_2 \alpha = 0.
\ee 
Thus, we get:
\be\label{phi-derivative}
\frac{\partial \phi}{\partial \Psi} = 
\frac{\phi_\alpha}{2} (\tlX_-\otimes \tlX_- + \tlX_+\otimes \tlX_+ )+
\frac{\phi_\beta}{6} ( \tlX_+ \otimes \tlX_- + \tlX_-\otimes \tlX_+ 
+ 4 \tlX\otimes \tlX).
\ee
where $\phi_\beta,\phi_\alpha$ are the partial derivatives of the function 
$\phi=\phi(\alpha,\beta)$ viewed as a function of $\alpha,\beta$.

\bigskip
\noindent{\bf Metricity equation.}
Having obtained a convenient representation (\ref{phi-derivative})
for the derivative of the
function $\phi$ with respect to the curvature field $\Psi$ we are
ready to write the metricity equations (\ref{metricity}) in a simple
form. Let us decompose the 2-form $B^i=B^{AB}$ into a basis of
rank 2 spinors:
\be
B = B_+ \tlX_+ + B_- \tlX_- + B \tlX,
\ee
where $B_{\pm}, B$ are some 2-forms. The trace of the product of two of these
2-forms is easy to compute:
\be
(B\wedge B) = 2 B_+ \wedge B_- - \frac{1}{2} B\wedge B.
\ee
Taking into account (\ref{phi-derivative}), after some algebra,
one obtains for the non-trivial part 
of the metricity equations (\ref{metricity}):
\be\label{metricity-1}
B_+ \wedge B_+ = B_- \wedge B_- = - \phi_\alpha 
\left( B_+ \wedge B_- - \frac{1}{4} B\wedge B \right), \\ \nonumber
B_+ \wedge B = B_- \wedge B = 0, \\ \nonumber
2 B_+ \wedge B_- + B\wedge B = - 2\phi_\beta
\left( B_+ \wedge B_- - \frac{1}{4} B\wedge B \right).
\ee
In the metric case (usual GR) the right hand side of all these equations is
zero. 

\bigskip
\noindent{\bf Interpretation.} 
Let us first discuss how the equations (\ref{metricity-1}) should be interpreted. These
equations were obtained by varying the action with respect to field $\Psi$. As such,
they should be interpreted as equations on $\Psi$, not on $B$. In the case of usual
Pleba\'nski gravity these equations are independent of $\Psi$, and this interpretation
is impossible. However, in the case considered the right hand side of these equations
is a (non-trivial) function of $\Psi$, and so they can be used to solve for components
of $\Psi$ in terms of components of $B$, at least in principle. 

In order to obtain such relations between $B$ and $\Psi$ in a more explicit form
it is desirable to be able to characterise the 2-forms $B^i$ by some scalar functions.
These functions could then be related algebraically to components of $\Psi$ via
the ``metricity'' equations (\ref{metricity-1}). To this end it is desirable to choose
some basis of one-forms, which can then be used to decompose the 2-forms $B_\pm, B$. Functions
appearing as coefficients in this decomposition could then be used as ``components'' of $B_\pm, B$,
and related to the components of $\Psi$.

It is in the process of constructing such a decomposition of $B$ into some basic
two-forms that the notion of a metric appears. The idea is as follows. The 2-forms $B_\pm, B$ 
span (generically) a 3-dimensional subspace in the 6-dimensional space of (complex) 2-forms. Let us
denote this subspace by $\cS$. The idea is to declare $\cS$ to be the subspace
of self-dual 2-forms. The notion of self-duality depends on a metric. As we shall
soon see, declaring $\cS$ to be the subspace of self-dual 2-forms determines
the metric with respect to which these forms are self-dual modulo conformal
factor. The above idea is, of course, motivated by the fact that in the usual Pleba\'nski GR case, the 
2-forms $B_\pm, B$ are self-dual with respect to the natural metric that arises.
As we shall see, this way of interpreting $B_\pm, B$ as self-dual 2-forms  of some metric
will be instrumental in selecting the scalar function that can later be related to the
components of $\Psi$.

To specify a triple of 2-forms $B_\pm, B$ one needs $3\times 6=18$ numbers. As one is
free to perform internal ${\rm SL}(2)$ rotations, 3 of these numbers are pure gauge, which
leaves one with 15 ``degrees of freedom'' contained in $B_\pm, B$. As we have described,
demanding that this triple of 2-forms is self-dual with respect to some metric
determines this metric, up to the ambiguity in the choice of the conformal factor,
see more on this below. To specify a metric in 4 spacetime
dimensions one needs 10 numbers. Thus, the triple of 2-forms $B_\pm, B$ carries
5 more (complex) ``degrees of freedom'' as compared to a metric. These 5 degrees of freedom
can be translated into 5 degrees of freedom of $\Psi$ via the metricity equations.
The above considerations quantify the difference between the usual metric theory of gravity,
with the main dynamical field being a metric, and the non-metric theory (\ref{action}), where the
main dynamical field is a triple of 2-forms.

\bigskip
\noindent{\bf Metric.}
Now, to see how a (complex) metric arises let us choose an arbitrary volume form $vol$, and using
this form define a $3\times 3$ symmetric matrix $h^{ij}$ via:
\be
B^i \wedge B^j = h^{ij} \, vol.
\ee
We note that in the usual metric case the metricity equations state that the 
matrix $h^{ij}$ is a multiple of the identity matrix. When eigenvectors of $h^{ij}$ are
all linearly independent the matrix $h^{ij}$ can be diagonalised by an ${\rm SO}(3,\C)$ transformation. The 
matrix of this transformation is $(a_1,a_2,a_3)$, where $a_I, I=1,2,3$ are the
eigenvectors of $h^{ij}$ normalized such that $a_I^i a_J^i = \delta_{IJ}$. 
When all the eigenvalues are distinct the eigenvectors $a_I^i$ are uniquely defined 
(modulo sign). Consider the following two-forms:
\be
B_I:= \frac{a_I^i}{\sqrt{\lambda_I}} B^i,
\ee
where both signs of the square root can be taken.
The three 2-forms so-defined satisfy:
\be
B_I \wedge B_J = \delta_{IJ} \, vol.
\ee
Modulo the coefficient on the right-hand-side, this is the relation satisfied by the two-forms $B^i$
in the metric case. Thus, the two-forms $B_I$ can be used to construct a metric. This metric is
subject to the same discrete ambiguities as in the usual metric case. An additional ambiguity not
present in the metric case is that of the conformal factor. Indeed, it is clear that the above
construction determines the metric only modulo conformal factor. To fix this ambiguity one has to
fix the volume form to be used. A natural choice seems to be:
\be\label{volume-form}
vol= \frac{1}{3} B^i\wedge B^i,
\ee
which would make the two-forms $B_I$ satisfy exactly the same metricity equation (\ref{metr-usual}) 
as in the case of Pleba\'nski theory. However, other choices are possible, see more on this below.

\bigskip
\noindent{\bf Reality conditions.} In order for the metric defined above to be real Lorentzian
the two-forms $B^i$ have to satisfy certain reality conditions. To see what these conditions
are we recall the reality conditions that must be satisfied by $B^i$ in the usual metric case. 
These are discussed in some length in \cite{CDJ}, so we simply state them here. They read:
\be\label{reality}
B^i \wedge (B^i)^* = 0,
\ee
where the star denotes the complex conjugation,
plus the condition that the volume form (\ref{volume-form}) is real. Since in the
general case the metric is constructed from two-forms $B_I$ in exactly the same
way as one proceeds in Pleba\'nski theory, to guarantee that a real Lorentzian metric arises 
one must impose exactly the same reality conditions on $B_I$. However, since we have
assumed that the eigenvectors of $h^{ij}$ are all linearly independent, the two-forms
$B_I$ and $B^i$ span the same three-dimensional subspace in the space of two-forms.
The conditions on $B_I$ then simply state that this space must be orthogonal to the
complex conjugate space. Thus, the reality conditions can be imposed directly on the
forms $B^i$ and take exactly the same form as in the case of Pleba\'nski theory.

The reader should note, however, that previous to a discussion of how matter 
couples to this theory of gravity it is unclear what the correct reality conditions are. 
Indeed, the matter is anticipated to couple directly to the $B$ field, and it is not
clear why the metric constructed from $B$ -- rather secondary object -- should be
real. This can only be understood after matter propagation in the background of 
a general $B$ is considered, something we hope to return to in another paper of this
series.

\bigskip
\noindent{\bf Metricity equations explicitly.} Our idea is thus to declare the subspace $\cS$
spanned by $B_\pm, B$ to be the space of self-dual 2-forms of a certain metric. To this end, 
let us choose a basis of one-forms, which we will denote by $l,n,m,\mb$, so that the space
$\cS$ coincides with the one spanned by the triple
\be\label{sd-forms}
m\wedge l, \qquad n\wedge \mb, \qquad l\wedge n - m\wedge\mb.
\ee
These are exactly the two-forms self-dual with respect to the metric 
\be\label{metric}
ds^2 = 2l\otimes n - 2m\otimes\mb,
\ee 
so finding the metric according
to the above construction one also finds the basis $l,n,m,\mb$. This
basis is, of course, only defined modulo ${\rm SL}(2)$ rotations. Reality
conditions imposed on $B^i$ (or, equivalently, $B_I$, see above) then
translate into the condition that $l,n$ are real and $m^*=\mb$.

Once the basis $l,n,m,\mb$ is found, one can use the available freedom in 
choosing $l,n,m,\mb$ to make sure that the 2-forms $B_\pm$ have decompositions of the form:
\be\label{B-pm}
B_+ = m\wedge l + a n\wedge \mb, \qquad B_- = n\wedge\mb + a' m\wedge l,
\ee
which is demonstrated in the Appendix.
One can now substitute these expressions for the forms $B_\pm$, as well as a general
decomposition for the form $B$ into the metricity equations (\ref{metricity-1})
to find that, in the special spinor basis chosen (in which $\Psi_1=\Psi_3=0$ and $\Psi_0=\Psi_4$), 
the field $B$ has the following form:
\be\label{B*}
B = \tlX_-(m\wedge l + a\, n\wedge \mb) + \tlX_+(n\wedge \mb + a\, m\wedge l) 
+ \tlX c(l\wedge n- m\wedge \mb),
\ee
with $a,c$ related to $\phi_\alpha,\phi_\beta$ as follows:
\be\label{metricity-rels}
2a+ \phi_\alpha(1+a^2 + c^2/2) = 0, \qquad 
(1+a^2-c^2) + \phi_\beta(1+a^2 + c^2/2) = 0.
\ee
These equations should be interpreted as equations for the functions $\alpha,\beta$ in terms
of the functions $a,c$ the later appearing as components of the field $B$. Note that the fact
that only two functions both in $B$ and in $\Psi$ survived is related to the fact that
we have chosen a very special basis in the spinor space. The choice of this basis encodes
the remaining 3 degrees of freedom of both $B$ and $\Psi$. One could have repeated the
analysis similar to that above working in the general spinor basis. In that case the 
2-forms $B_\pm$ would be still given by expressions (\ref{B-pm}), and the 2-form $B$ would
have all 3 components, with 3 functions, say $c_\pm, c$ appearing in front of the 2-forms 
(\ref{sd-forms}). The 5 functions $a,a',c_\pm,c$ would then be related to 5 components
of the field $\Psi$ via the metricity equations. The corresponding formulae are rather
cumbersome, and we will refrain from giving them here. The approach that does not
rely on a special choice of the spinor basis may be convenient for some developments,
but we will not follow it in the present paper.
 
Let us note that when $\phi=0$ the solution to (\ref{metricity-rels})
is $a=0,c=1$, which, when substituted into (\ref{B*}),
gives the usual ``metric'' 2-form: 
\be\label{B-metric}
B= \tlX_+ m\wedge l+\tlX_- n\wedge\mb +\tlX(l\wedge n- m\wedge \mb).
\ee 

\bigskip
\noindent{\bf Different possible metric structures.}
As we have seen above, the fundamental field of the theory, the $B$ field, defines
a metric structure, only modulo ambiguity in the choice of the conformal factor. 
There are at least a few natural
metrics in the same conformal class that can be constructed from $B$. One of
such metrics is the one known as Urbantke metric  \cite{urbantke}. The
corresponding expression reads:
\be\label{urbantke}
\sqrt{|g^{\rm U}|}\, g^{\rm U}_{\mu\nu} = \frac13
\epsilon^{\alpha\beta\gamma\delta} {\rm Tr} \left( B_{\mu\alpha} B_{\beta\gamma}
B_{\delta\nu} \right) \, .
\ee
As it is not hard to check, this metric constructed directly from $B$, is in 
the same conformal class as the one introduced earlier via (\ref{metric}) based
on the choice of the tetrad $l,n,m,\mb$ that leads to say (\ref{B*}). However,
the conformal factor relating the two metrics is quite non-trivial.

Another natural possibility, already advocated above, is to require that the
volume form as defined by the metric coincides with the volume form
\be
\frac{1}{3} {\rm Tr} \left(B \wedge B \right).
\ee
This leads to yet a different possible mechanism for fixing the metric. Note that in 
the usual metric case when $B$ is of the form (\ref{B-metric}) all the above choices
of the conformal factor agree and there is no ambiguity in what the metric is. 

\bigskip
\noindent{\bf $B$ as a map.}
Above we have seen that the difference between the usual metric case
(\ref{B-metric}) and the general one (\ref{B*}) is due to appearance
of 5 more dynamical fields as compared to those described
by a metric. The components of $\Psi$ can be expressed in terms of these
degrees of freedom via the metricity equations. Note however that in spite
of hiving introduced more degrees of freedom into the theory the 
number of physical degrees of freedom has not changed --- the theory
(\ref{action}) still propagates 2 polarisations of the graviton. This
e.g. follows from the analysis of the structure of the phase space of
the theory performed in \cite{Bengtsson:2007zx}. It is important to
understand the geometrical interpretation of the new scalar fields
introduced. One viewpoint that will prove quite instrumental below is to
regard $B$ as a map from the $\su(2)$ Lie algebra spanned by $\tlX_\pm,\tlX$ to 
the space of self-dual 2-forms. This map thus defines the notion of self-dual
2-forms, and thus defines a metric with respect to which these two-forms
are self-dual. However, this map contains more information than just that
about a metric, which, for example, follows from the counting of degrees of freedom
contained in $B$ that we performed above. In the metric case the field
$B$ carries information just about a metric, and can be identified with
the identity map after certain additional techniques are developed. The
general non-metric $B$ corresponds to a map that is not an identity.

\bigskip
\noindent{\bf Type III case.}
When $\Psi$ is Petrov type III the above analysis is not applicable. In this
case the spinor basis can be chosen so that the field $\Psi$ becomes 
\be
\Psi=\alpha (\tlX_+ \otimes \tlX + \tlX\otimes \tlX_+).
\ee
We will also need the expression for its square:
\be
\Psi^2 = -\frac{\alpha^2}{2} \tlX_+ \otimes \tlX_+.
\ee
Note that this is traceless, so no need to subtract the trace part. 

The metricity equations become:
\be
B_-\wedge B_- = B\wedge B_- = 2B_+\wedge B_- + B\wedge B = 0, \\ \nonumber
B_+ \wedge B \sim \alpha (B_+\wedge B_- - (1/4) B\wedge B), \qquad
B_+\wedge B_+ \sim \alpha^2 (B_+\wedge B_- - (1/4) B\wedge B).
\ee
A solution of these equations is:
\be
B= \tlX_-(m\wedge l) + \tlX_+(n\wedge \mb + a(l\wedge n-m\wedge\mb)) + 
\tlX(l\wedge n-m\wedge\mb),
\ee
where $a$ is proportional to $\alpha$. Thus, a ``canonical''
quadruple of one-forms appears even in the type III case, and the field $B$ 
is still purely self-dual. We will not consider the Petrov type III case 
further in the present paper.

\section{Spinor techniques}
\label{sec:spinors}

So far we only dealt with algebraic equations for $B$ involving $\su(2)$
matrices, and we saw that these are much easier to solve if one
employs spinor notations. The spinors we have introduced above are
``internal'' ones, in that they simply replace the internal $\su(2)$ 
indices by a pair of spinor ones. In this section we will introduce 
different spinors, the ones that will allow us to simply computations
with forms and vector fields considerably. In usual gravity, the
``internal'' and ``spacetime'' spinors coincide, and this statement
is equivalent to the statement that $B$ has the form (\ref{B-metric}).
In other words, after spinor methods are employed, $B$ of the metric
case is the identity map that identifies the ``internal'' and ``spacetime'' spinors.
In our more general case we will need to introduce two different 
types of spinors, one for dealing with ``internal'' indices,
one for spacetime indices. In this section we will remind the reader
how the spacetime indices can be converted into spinor ones, and how
this simplifies certain computations. The material reviewed here is
standard, see e.g. \cite{Penrose}. The only non-standard point
is that the spinors we are going to deal with in this section are 
different from the ones considered above. For this reason, we have denoted the 
``internal'' spinors by tilded letters and reserved the usual
notation for the ``usual'' spacetime spinors. The corresponding ``spacetime''
spinor basis will be denoted by $i^A, o^A$. 

There are several different ways to introduce spinor techniques. We will
first present what we think is a more geometric way, and then give
the usual one of e.g. \cite{Penrose}.

\bigskip
\noindent{\bf Two-forms as maps.}
It will be extremely useful to think of the 2-forms from space
$\cS$ as maps acting on one-forms, and sending them to vector fields. 
This way of thinking about 2-forms will allow us to develop a 
spinor approach to forms. What we
are about to describe is standard in the usual GR. 

Let us consider a given 2-form $B$. Any one-form $k$ can be exterior multiplied
by $B$ with the result being a 3-form $k\wedge B$. Now, given a basis 
$l,n,m,\mb$ in the space of one-forms we can use this basis to produce 
4 numbers out of the 3-form $k\wedge B$. Indeed, multiplying $k\wedge B$
(from say the right) by each of the basis one-forms one obtains a 4-form
that must be proportional to the ``volume'' 4-form $l\wedge n\wedge m\wedge\mb$.
Let us introduce a special notation for this proportionality coefficient. Thus,
we write:
\be
k\wedge B \wedge e^I := (k\wedge B \wedge e^I) l\wedge n\wedge m\wedge\mb,
\ee
which defines the quantity $(k\wedge B \wedge e^I)$. Here $e^I$ is any
of the basis one-forms. One can use the 4 quantities $(k\wedge B \wedge e^I)$
to form a vector field $v^a:=\sum_I (k\wedge B \wedge e^I) e^{Ia}$. Here
$e^{Ia}$ denotes the basis of vector fields dual to $e^I=e^I_a$ one forms.
By definition: $l^a n_b + n^a l_b - m^a \mb_b - \mb^a m_b = \delta^a_b$, 
where $\delta^a_b$ is the Kronecker delta - a mixed tensor of rank 2, 
and $l^a n_a = 1, m^a \mb_a=-1$, while all other scalar products are zero.

The above discussion allows us to interpret any 2-form $B$ as a map from
the space of one-forms to the space of vector fields. It is
useful to work out this map for the self-dual 2-forms forming a basis in $\cS$.
An elementary computation gives:
\be\label{two-form-maps}
m\wedge l: \,\,\lower1ex\vbox{\hbox{$n_a\to -m^a$}\hbox{$\mb_a\to -l^a$}}, \qquad
n\wedge \mb: \,\,\lower1ex\vbox{\hbox{$l_a\to \mb^a$}\hbox{$m_a\to n^a$}}, \qquad
l\wedge n - m\wedge\mb: \,\,
\lower4ex\vbox{\hbox{$l_a\to -l^a$}\hbox{$n_a \to n^a$}
\hbox{$m_a\to -m^a$}\hbox{$\mb_a\to \mb^a$}}
\ee
A similar map for the anti-self-dual forms is given by:
\be\label{two-form-maps-anti}
l\wedge \mb: \,\,\lower1ex\vbox{\hbox{$n_a\to -\mb^a$}\hbox{$m_a\to -l^a$}}, \qquad
n\wedge m: \,\,\lower1ex\vbox{\hbox{$l_a\to -m^a$}\hbox{$\mb_a\to -n^a$}}, \qquad
l\wedge n + m\wedge\mb: \,\,
\lower4ex\vbox{\hbox{$l_a\to l^a$}\hbox{$n_a \to -n^a$}
\hbox{$m_a\to -m^a$}\hbox{$\mb_a\to \mb^a$}}
\ee

\bigskip
\noindent{\bf Spinor interpretation.}
The above described way of viewing two-forms as maps from one-forms to
vector fields will allow
us to develop a completely form-free approach, in which forms will get 
replaced by spinorial objects. Let us first deal with the basic one-forms
(tetrad-forming) $l,n,m,\mb$. To replace them by spinor 
quantities we will have to introduce
the so-called primed spinors, in addition to the unprimed spinors we have
considered so far. Primed spinors form a different representation of the
Lorentz group. The vector representation is obtained by tensoring together
the primed and unprimed representations. This is the reason why the primed
spinors are necessary to describe one-forms. 

The primed spinors are denoted by a symbol with a primed spinorial index next to 
it, e.g. $\lambda^{A'}$ is a rank one primed spinor. Primed indices are
similarly raised and lowered with their corresponding $\epsilon^{A'B'}$
anti-symmetric rank 2-spinor. The raising-lowering conventions are
the same (\ref{raise-lower}) as for the unprimed spinors. It is convenient
to introduce a basis in the space of primed spinors. We get the primed
spinor basis $i^{A'}, o^{A'}$. 

Let us now introduce an important set of relations which are to provide the
identification between rank 2 mixed primed-unprimed spinors and one-forms.
We write:
\be\label{basis-spinor}
l_a = o_A o_{A'}, \qquad n_a = i_{A} i_{A'}, \qquad 
m_a = o_A i_{A'}, \qquad \mb_a = i_A o_{A'}.
\ee
As it is clear, these relations depend both on the spinor basis chosen as 
well as on the basis in the space of one-forms. These relations can be
combined into a single quantity
\be
\theta_a^{AA'} := l_a i^A i^{A'} + n_a o^A o^{A'} - m_a i^A o^{A'} - \mb_a o^A i^{A'},
\ee
which can be viewed as a map from rank 2 mixed spinors into one-forms, the map
precisely given by (\ref{basis-spinor}).

Using the above relations we can give a description of self-dual 2-forms in terms
of symmetric rank 2 unprimed spinors. Indeed, let us identify:
\be\label{self-dual}
m\wedge l = X_+, \qquad n\wedge \mb = X_-, 
\qquad l\wedge n-m\wedge \mb = -2X
\ee
for the self-dual 2-forms and
\be\label{anti-self-dual}
l\wedge \mb = X_+', \qquad n\wedge m = - X_-', 
\qquad l\wedge n+m\wedge \mb = 2X'
\ee
for the anti-self-dual ones. Here the unprimed rank 2 spinors are
essentially the same to the ones we already encountered in (\ref{tl-X}),
and the primed spinors are similar:
\be\label{X}
i^{A} i^{B} := X_-^{AB}, \qquad 
o^{A} o^{B} := X_+^{AB}, \qquad
\frac{1}{2} (i^{A} o^{B} + o^{A} i^{B} ) = X^{AB}, \\ \nonumber
i^{A'} i^{B'} := {X_-'}^{AB}, \qquad 
o^{A'} o^{B'} := {X_+'}^{AB}, \qquad
\frac{1}{2} (i^{A'} o^{B'} + o^{A'} i^{B'} ) = {X'}^{AB}.
\ee 
It is then easy to check that the action of the rank 2 spinors on 
the mixed spinors (\ref{basis-spinor}) is exactly as in
(\ref{two-form-maps}). Indeed, let us consider e.g. $(m\wedge l)(n)$.
The corresponding spinor quantity is $X_+^{AB} i_B i^{A'} 
= - o^A i^{A'}= -m^a$, as in (\ref{two-form-maps}). Similarly,
$(l\wedge\mb)(n)$ is given by ${X_+'}^{A'B'} i_{B'} i^A = -
i^A o^{A'} = -\mb^a$ as in (\ref{two-form-maps-anti}).

\bigskip
\noindent{\bf Another way to describe two-forms as rank 2-spinors.}
The following is a more standard approach to spinors, see e.g.
\cite{Penrose}. One starts by postulating the relations
(\ref{basis-spinor}) between the 4 basic one-forms and spinors
of the spinor basis. This relation allows one to replace any spacetime
index by a mixed pair of spinor indices. It is then easy to see
that e.g. self-dual two forms becomes rank 2 unprimed spinors. Indeed,
let us consider a 2 form $B=v\wedge w$ obtained by an exterior product of 
two one-forms $v,w$. Introducing indices this reads: $B_{ab} = 2v_{[a} w_{b]}$.
The last quantity here is given by:
\be
v_{[a} w_{b]} = v_{[AA'} w_{BB']} = -
\frac{1}{2} (v\wedge w)_{AB} \epsilon_{A'B'} + \frac{1}{2}
(v\wedge w)_{A'B'} \epsilon_{AB},
\ee
where the quantities $(v\wedge w)_{AB}, (v\wedge w)_{A'B'}$ are
defined as:
\be
(v\wedge w)_{AB} = v_{(A}^{A'} w_{B)A'}, \qquad 
(v\wedge w)_{A'B'} = - v_{(A'}^A w_{A|B')}.
\ee
The choice of minus signs in these formulae is so that they agree with
(\ref{self-dual}), (\ref{anti-self-dual}).

\bigskip
\noindent{\bf $B$ as a rank 4 spinor.}
The above description of self-dual two-forms as rank 2 unprimed symmetric
spinors allows us to interpret the $B$ field (\ref{B*}) as a rank 4
unprimed spinor. Indeed, using (\ref{self-dual}) we write:
\be\label{b-spinor}
B = a(\tilde{X}_-\otimes X_- + \tilde{X}_+ \otimes X_+) + 
\tilde{X}_-\otimes X_+ + \tilde{X}_+\otimes X_- - 2c \tilde{X}\otimes X.
\ee
We see, therefore, that the $B$ field interpreted as a map is the identity
map (\ref{id}) in the metric case $a=0,c=1$ (in this case the ``internal''
and ``spacetime'' spinors coincide, as we shall see below), and deviates from 
identity in the non-metric case. This gives one way to understand
what the non-metric deformation does to $B$. For completeness, let
us also give the expression for $B$ as a rank 4 spinor in the type III case: 
\be
B= -2 a \tilde{X}_+ \otimes X + Id.
\ee

\bigskip
\noindent{\bf Metric compatible derivative operator.}
Having a basis of one-forms $l,n,m,\mb$ one has the metric (\ref{metric}), and can consider
the derivative operator compatible with this metric. We will denote this
operator by $\nabla$. Spinor representation $\nabla_{AA'}$ of this derivative
operator will be very convenient for us in what follows. In order to describe
the action of this operator on forms, we just have to understand its action
on basic spinors, as we have already translated one- and two-forms that
are of interest for us into spinor form. Thus, we introduce what is known 
as {\it spin-coefficients} according to the following formulae:
\be
\gamma^+_{AA'}: = - i^{B} \nabla_{AA'} i_B, \qquad 
\gamma^-_{AA'}: = - o^{B} \nabla_{AA'} o_B, \qquad 
\frac{1}{2} \gamma_{AA'} := i^{B} \nabla_{AA'} o_B =
o^{B} \nabla_{AA'} i_B.
\ee
The derivatives of the basic spinors are then given by the following
formulae:
\be\label{der-basis}
\nabla_{AA'} i_B = -\frac{1}{2} \gamma_{AA'} i_B - \gamma^+_{AA'} o_B, 
\qquad 
\nabla_{AA'} o_B = \gamma^-_{AA'} i_B + \frac{1}{2} \gamma_{AA'} o_B.
\ee
Let us note that the anti-symmetric tensor $\epsilon_{AB}=o_A i_B -i_A o_B$
that serves the role of the metric in the spinor space is preserved by the 
derivative operator $\nabla: \nabla\epsilon_{AB}=0$, which is 
quite easy to check using (\ref{der-basis}). This fact is
related to the fact that $\nabla$ preserves the metric constructed from 
the basis of one-forms. 

Using (\ref{der-basis}) it is easy to work out
the action of the derivative operator $\nabla$ on the basic two-forms.
We will only need these results for the self-dual 2-forms which 
appear in the decomposition of $B$. Thus, using (\ref{self-dual})
and (\ref{der-basis}) we get:
\be\label{der-basis-2}
\nabla_{AA'} X_-^{BC} = -\gamma_{AA'} X_-^{BC} - 2\gamma_{AA'}^+ X^{BC}, \qquad
\nabla_{AA'} X_+^{BC} = \gamma_{AA'} X_+^{BC} + 2\gamma_{AA'}^- X^{BC}, 
\\ \nonumber
\nabla_{AA'} X^{BC} = \gamma_{AA'}^- X_-^{BC} - \gamma_{AA'}^+ X_+^{BC}.
\ee
We now note that these derivatives of self-dual 2-forms 
can be converted into vector fields. This is done simply by contracting
the unprimed index of $\nabla_{AA'}$ with an unprimed index of the
tensors $X_\pm,X$. Thus, we introduce the vector fields 
$(\nabla X_\pm), (\nabla X)$ corresponding to
$\nabla X_\pm, \nabla X$, and get for the corresponding mixed spinors:
\be
(\nabla X_-)^{AA'} = - X_-^{AB}\gamma_{B}^{A'} - 2 X^{AB} \gamma_{B}^{+A'}, \qquad
(\nabla X_+)^{AA'} = X_+^{AB}\gamma_{B}^{A'} + 2 X^{AB} \gamma_{B}^{-A'},
\\ \nonumber
(\nabla X)^{AA'} = X_-^{AB} \gamma_{B}^{-A'} -  X_+^{AB}\gamma_{B}^{+A'}.
\ee
These formulae can be rewritten much more compactly by
suppressing the spinor indices. Thus, implying matrix multiplication, we can re-write:
\be\label{der-x}
(\nabla X_-) = - X_- \gamma  - 2 X \gamma^+, \qquad
(\nabla X_+) = X_+ \gamma  + 2 X \gamma^-, \qquad
(\nabla X) = X_- \gamma^- -  X_+\gamma^+.
\ee

\section{Compatibility equations}
\label{sec:comp}

Having understood what the modified ``metricity'' equations imply, and 
developed spinor techniques for dealing with forms and their derivatives, we are
ready to analyse another set of equations that follow from the action
(\ref{action}) - the compatibility equations between the 2-form $B$ and
the connection $A$. As we shall see in this section, these equations 
allow one to find the connection $A$ in terms of the derivatives of
the quantities that appear in the expression (\ref{B*}) for $B$.
Spinor techniques developed in the previous section will be of great
help here.

\bigskip
\noindent{\bf The compatibility condition.}
The condition relating $B$ and $A$ is obtained by varying the action
(\ref{action}) with respect to the connection $A$. It reads $\cD B^i=0$,
where $\cD$ is the covariant derivative with respect to the connection $A$.
In spinorial notations this equation reads $dB+[A,B]=0$, or, with indices:
\be\label{comp-eq}
dB^{AB}+ A^{AC}\wedge B_C^B - B^{AC}\wedge A_C^B = 0.
\ee
To rewrite this in a more manageable form let us decompose the
connection into the basis of ``internal'' rank 2 spinors:
\be
A= \tlX_- A_- + \tlX_+ A_+ +\tlX A.
\ee
The commutator present in (\ref{comp-eq}) is then easy to compute:
\be\label{comm}
[A,B]=\tlX_-(A_-\wedge B - A\wedge B_-) + \tlX_+(A\wedge B_+ - A_+\wedge B)
+ 2\tlX(A_-\wedge B_+ - A_+\wedge B_-).
\ee

\bigskip
\noindent{\bf Derivatives of the basis spinors.}
To compute the derivative $dB$ we need to act on the spinors $\tli^A,\tlo^A$. 
We could have chosen a constant basis such that the basis spinors
do not depend on a point in spacetime. However, to write the field $\Psi$
is its most convenient for computations form we have chosen to adapt
the spinor basis at every point to the field $\Psi$ and its
proper spinors. Thus, the basis in which the field $\Psi$ has the
simple form (\ref{psi-gauge}) is generally not constant. Thus, we have
to allow for no-vanishing derivatives of $\tli^A, \tlo^A$. Let us proceed
similarly to what was done in the previous section and decompose
these derivatives into the basic spinors:
\be\label{der-spinors}
d\tli^A := h \tli^A + g\tlo^A, \qquad d\tlo^A := f\tli^A - h\tlo^A.
\ee
Here, $h,g,f$ are one-forms, and 
to write the second relation we have used the normalization 
condition $\tli^A \tlo_A=1$, which implies that the one-form coefficient
in front of $\tlo^A$ in the second relation is minus the one-form coefficient
in front of $\tli^A$ in the first. 

\bigskip
\noindent{\bf Gauge transformations.}
It is useful to discuss what gauge transformations that act on the 
basis $\tli^A,\tlo^A$ translate to when they act on the one-forms $h,g,f$. Choosing directions of
the spinors $\tli^A,\tlo^A$ and requiring them to be normalized does not fix them
completely. Indeed, there is still a freedom of rescaling: $\tli^a\to \alpha \tli^A,
\tlo^A\to (1/\alpha) \tlo^A$, 
where $\alpha$ is a complex number different from zero.
It is easy to see that under this gauge transformation:
\be
h\to h+ \alpha^{-1} d\alpha, \qquad g\to \alpha^2 g, \qquad f\to \alpha^{-2} f,
\ee
and so $h$ transforms as a ${\rm U}(1)$ connection, while the other one-forms
$g,f$ transform as Higgs fields.

\bigskip
\noindent{\bf Derivatives of the basis rank 2 spinors.}
Using (\ref{der-spinors}) it is easy to compute the derivatives of the
basic rank 2 spinors $\tlX_\pm, \tlX$. We get:
\be\label{der-rank-2}
d\tlX_- = 2h \tlX_- + 2g \tlX, \qquad d\tlX_+=2f \tlX - 2h \tlX_+, 
\qquad d\tlX = f\tlX_- + g\tlX_+.
\ee
These formulae are quite similar to (\ref{der-basis-2}) except that
``internal'' spinors are considered.

\bigskip
\noindent{\bf The compatibility equations.}
Combining (\ref{comm}) and (\ref{der-rank-2}) we can write down the
compatibility equations that follow from (\ref{comp-eq}):
\be\nonumber
2h\wedge B_- + dB_- + f\wedge B + A_-\wedge B - A\wedge B_- = 0, \\
\label{comp*}
g\wedge B + dB_+ - 2h\wedge B_+ + A\wedge B_+ - A_+\wedge B =0, \\
\nonumber
2g\wedge B_- + dB + 2f\wedge B_+ + 2(A_-\wedge B_+ - A_+\wedge B_-)=0.
\ee
Each of these equations is a condition that a certain 3-form vanishes. Therefore,
each gives rise to 4 equations when projected onto a basis of 3-forms.
Thus, the compatibility equations are $3\times 4=12$ equations for 3 one-forms
$A_\pm,A$. We see that the number of equations matches the number of
unknowns. Below we will write down an explicit solution for $A$ in 
terms of the $B$ field of the form (\ref{B*}).

\bigskip
\noindent{\bf Solving the compatibility equations.}
To solve the equations (\ref{comp*}) we decompose the one-forms
$A_\pm,A$ into the same basis $l,n,m,\mb$ that was used to write
down the expression (\ref{B*}) for $B$. We then multiply each of the 
equations (\ref{comp*}) by one-forms $l,n,m,\mb$ in turn, and
thus extract components of these equations. This procedure allows
us to find $A$ in terms of $B$. 

A very convenient way of doing this is to use the spinor method 
developed in the previous section. In view of (\ref{basis-spinor})
decomposing $A_\pm,A$ into basic one-forms is equivalent to
introducing the rank 2 mixed spinors $A_{\pm AA'}, A_{AA'}$. The field
$B$ can in turn be represented by a rank 4 unprimed spinor 
(\ref{b-spinor}), or, equivalently, each of the 2-forms $B_\pm,B$ can
be represented by a rank 2 unprimed spinor. We have:
\be\label{b1}
B_-=X_+ + a X_-, \qquad B_+ = X_- + aX_+, \qquad B = -2c X,
\ee
where we have omitted spinor indices for compactness. Note that
it is the ``spacetime'' rank 2 spinors $X_\pm,X$ 
that are used here. These should not be confused with the ``internal''
ones. To solve compatibility conditions we will need the inverses of these
matrices, where the inverse $B^{-1}$ of $B$ is defined so that
$(B^{-1})^{AC} B_C^B = \epsilon^{AB}$. These are easy to find, 
we get:
\be\label{b-x}
B_-^{-1} = - X_- - (1/a) X_+, \qquad B_+^{-1} = - X_+ - (1/a) X_-,
\qquad B^{-1} = - (2/c) X.
\ee

In preparation for solving the equations (\ref{comp*}), let us introduce
special notations for the commutators that appear in (\ref{comp*}):
\be
C_- = A_-\wedge B - A\wedge B_-, \qquad 
C_+=A\wedge B_+ - A_+\wedge B , \qquad
C= A_-\wedge B_+ - A_+\wedge B_-.
\ee
Using the spinorial notations for all the quantities, and converting
the 3-forms that appear here into vector fields, these expressions
can be rewritten as:
\be
C_-^{FF'}=B^{FE}A_{-E}^{F'} - B_-^{FE} A_E^{F'}, \qquad
C_+^{FF'}=B_+^{FE}A_{E}^{F'} - B^{FE} A_{+E}^{F'}, \qquad
C^{FF'}=B_+^{FE}A_{-E}^{F'} - B_-^{FE} A_{+E}^{F'},
\ee
or, in a more compact form, implying matrix multiplication:
\be\label{c-ab}
C_-=BA_{-} - B_- A, \qquad
C_+=B_+A - B A_{+}, \qquad
C=B_+A_{-} - B_- A_{+}.
\ee
Now, assuming $C_\pm,C$ are known, it is easy to solve for $A_\pm, A$. Thus,
for example, multiplying the first equation by $B_+ B^{-1}$ and the second by
$B_- B^{-1}$, adding the results and then subtracting the third equation we 
get:
\be
(B_+ B^{-1} B_- - B_- B^{-1} B_+)A = C - B_+ B^{-1} C_- - B_- B^{-1} C_+.
\ee
One then checks that
\be
B_+ B^{-1} B_- - B_- B^{-1} B_+ = \frac{1-a^2}{c} {\rm Id},
\ee
and therefore
\be\label{a}
A = \frac{c}{1-a^2} \left( C - B_+ B^{-1} C_- - B_- B^{-1} C_+\right).
\ee
By similar manipulations
\be\label{a-pm}
A_- = \frac{a}{c(1-a^2)} \left( -B B_-^{-1} C + B_+ B_-^{-1} C_- + C_+ \right),
\quad
A_+ = \frac{a}{c(1-a^2)} \left( -B B_+^{-1} C + C_- + B_- B_+^{-1} C_+ \right).
\ee

\bigskip
\noindent{\bf Dependence on one-forms $h,g,f$.}
One should now simply take the expressions for $C_\pm,C$ from
the compatibility conditions (\ref{comp*}) and substitute these
into the above expressions to find $A$. It can now be checked that 
the one-forms $h,g,f$ appear in the connection $A$ in the following
simple way:
\be\label{tilde-a}
A = - \tlX_- f + \tlX_+ g + 2\tlX h + \tilde{A},
\ee
where $\tilde{A}$ is independent of $h,g,f$. This is as expected, for
\be
 - \tlX_- f + \tlX_+ g + 2\tlX h 
= \left( \begin{array}{cc} h & g \\ f & -h \end{array}
\right) = dG^{-1} \cdot G, 
\ee
where $G$ is an ${\rm SL}(2)$ transformation that sends the basis $\tli^A,\tlo^A$
into some constant basis. 

We note that the components of $\tilde{A}$ have a very simple meaning:
\be\label{ta}
\tli^A \cD \tli_A = - \tilde{A}_+, \qquad \tlo^A \cD \tlo_A = - \tilde{A}_-, 
\qquad \tlo^A \cD \tli_A = \tli^A \cD \tlo_A = \frac{1}{2} \tilde{A},
\ee
where $\cD$ is the covariant derivative with respect to the connection $A$. 
These relations are easy to check using (\ref{tilde-a}). Thus, the 
components of $\tilde{A}$ are the spin-coefficients, but for the
``internal'' spinor basis instead of the ``spacetime'' one. As we shall
see very soon, in the metric case the two sets of spin-coefficients
coincide. However, in a more general non-metric situation the two
sets are different. Correspondingly, the covariant derivatives that
act on ``internal'' and ``spacetime'' spinor indices are different.

\bigskip
\noindent{\bf Spin-coefficients.}
To compute $\tilde{A}$ we have to substitute
\be\label{nB}
\tilde{C}_- = -(\nabla B_-), \qquad \tilde{C}_+ = -(\nabla B_+), 
\qquad \tilde{C} = -(1/2) (\nabla B),
\ee
where $B_\pm,B$ are given by (\ref{b1}) into the formulae 
(\ref{a-pm}), (\ref{a}). Note that we are free to replace the
usual derivatives here by the covariant ones, as their (exterior product) 
action on forms is the same. Thus, using (\ref{der-x}) we have:
\be
\tilde{C}_- = X_-(a\gamma -\nabla a)- X_+ \gamma + 2X (a \gamma^+-\gamma^-),
\qquad 
\tilde{C}_+ = X_- \gamma - X_+(a\gamma + \nabla a) + 2X(\gamma^+-a\gamma^-),
\\ \nonumber
\tilde{C} = cX_- \gamma^- - cX_+ \gamma^+ + X\nabla c.
\ee
This gives:
\be\nonumber
\tilde{A}_- = \frac{1}{2c(1-a^2)}\Big( c(X_+ +aX_-)\nabla c -
2(X_- + aX_+)\nabla a + (1+c^2-a^2)\gamma^- - 2(1-c^2+3a^2) X\gamma^- 
\\ \nonumber
- (1-c^2-a^2) a\gamma^+ + 2a(3-c^2+a^2) X\gamma^+ + 4a(X_- - a X_+)\gamma \Big),
\\ 
\tilde{A}_+ = \frac{1}{2c(1-a^2)}\Big( c(X_- +aX_+)\nabla c -
2(X_+ + aX_-)\nabla a + (1+c^2-a^2)\gamma^+ + 2(1-c^2+3a^2) X\gamma^+ 
\\ \nonumber
- (1-c^2-a^2) a\gamma^- - 2a(3-c^2+a^2) X\gamma^- + 4a(a X_- - X_+)\gamma \Big),
\\ \nonumber
\tilde{A} = \frac{1}{1-a^2}\left(cX\nabla c - 2a X\nabla a + 
(1+a^2)\gamma + (1-c^2+a^2)(X_+ \gamma^+ - X_- \gamma^-) 
+2a(X_- \gamma_+ - X_+ \gamma^-) \right).
\ee
As a check of these results, let us note that in the metric case $a=0,c=1$
and the computed ``internal'' spin-coefficients coincide with the
``spacetime'' ones. 

For practical computations, e.g. in the case of spherical symmetry,
many of the quantities we have been discussing simplify, and the
spinor method is an overkill. A different method of computing the
spin-coefficients that avoids spinors and uses only exterior product of
forms is given in the Appendix.

\section{Field equations}
\label{sec:eqs}

In this section, using the results we have obtained above, we
write down the field equations for theory (\ref{action}).
As we shall see, these become equations for the derivatives of the one-forms 
$l,n,m,\mb$ as well as for the curvature functions $\alpha,\gamma$
and their derivatives.

\bigskip
\noindent{\bf The field equations.}
The equations we are to discuss in this section are obtained by varying
the action with respect to $B^i$. We get:
\be\label{eqs-f}
F^i + (\Lambda \delta^{ij} + \Psi^{ij} + \phi \delta^{ij}) B^j = 8\pi G T^i,
\ee
where the right hand side is obtained by varying the matter part of the
action with respect to $B^i$. Note that in the discussion above we
have tacitly assumed that matter only couples to the $B$ field and not
to $A, \Psi$. This is true for the gauge fields, Maxwell field in particular,
which will serve as a principal example of test matter in the subsequent
papers of the series.

\bigskip
\noindent{\bf Spinor form.}
As before, it is useful to write these equations using spinors. We 
will make use of the gauge in which $\Psi$ has the form (\ref{psi-gauge})
and $B$ has the form (\ref{B*}). Let us first compute the matrix that
appears in front of $B^i$ in the second term of (\ref{eqs-f}):
\be
(\Lambda+\phi)\, {\rm Id} + \Psi = 
\alpha (\tlX_-\otimes \tlX_- + \tlX_+\otimes \tlX_+ ) + \\ \nonumber
(\beta + \Lambda + \phi) ( \tlX_+ \otimes \tlX_- + \tlX_-\otimes \tlX_+)
 + 2(2\beta-(\Lambda+\phi)) \tlX\otimes \tlX.
\ee
The second term on the left hand-side of (\ref{eqs-f}) 
can now be easily computed:
\be\label{psi-b}
((\Lambda+\phi) {\rm Id} + \Psi) B= 
\tilde{X}_- (\alpha B_+ +
(\beta+\Lambda+\phi) B_-) + \tilde{X}_+ (\alpha B_- +
(\beta+\Lambda+\phi) B_+) + (\Lambda+\phi-2\beta) \tilde{X} B.
\ee
To get the field equations it remains to split 
each of the 2-forms $F^{AB}, T^{AB}$ into their self-dual and 
anti-self-dual parts. Thus, the field equations
can be split as:
\be\label{eqs-1}
F' = 8\pi G T', \\ \label{eqs-2}
F + \tilde{X}_- (\alpha B_+ +
(\beta+\Lambda+\phi) B_-) + \tilde{X}_+ (\alpha B_- +
(\beta+\Lambda+\phi) B_+) + (\Lambda+\phi-2\beta) \tilde{X} B = 8\pi G T.
\ee
We note that for conformally invariant matter (e.g. Yang-Mills fields) 
the self-dual part of the stress-energy 2-form vanishes (this is equivalent
to the statement that the trace of the stress-energy tensor vanishes).
We thus see that field equations for theory (\ref{action}) have interpretation
quite similar to that in usual GR. The curvature 2-form $F^{AB}$ splits
into two parts: ``Weyl'' part $F$ and ``Ricci'' part $F'$. 
The first ``Weyl'' part gets related to the ``curvature field'' 
$\Psi$ components $\alpha,\gamma$, as well as to the trace part
of the stress-energy tensor, via equation (\ref{eqs-2}), while the second 
``Ricci'' part is proportional to the traceless part $T'$ of the stress-energy. 

Thus, modifications due to non-metricity 
are not so dramatic after all. The main difference with the usual GR case
is that the connection components (and thus curvature $F$ components) depend not
only on the frame $l,n,m,\mb$ and its derivatives but also on the other, non-metric
components of $B$ and their derivatives. The later can be expressed in terms of $\Psi$
components via the metricity equations, and it is in this form that the field
equations are most convenient for analysis. We thus get 
the schematic structure described by equation (\ref{intr-1}).

\bigskip
\noindent{\bf Curvature.}
The curvature of $A$ is easy to compute. We have:
\be
F_- = dA_- + A_- \wedge A, \qquad
F_+ = dA_+ + A \wedge A_+, \qquad
F = dA + 2 A_- \wedge A_+.
\ee
One should now substitute (\ref{tilde-a})
into these expressions. The result does not seem to be very illuminating, and we
shall refrain from giving it here.

\section{Bianchi identity}
\label{sec:bianchi}

In this section we derive Bianchi identities for theory (\ref{action})
and discuss the energy conservation. Spinor method is most effective
for this purpose, and will be used heavily in this section.

\bigskip
\noindent{\bf ``Bianchi'' identity.} A very important identity,
analogous to the Bianchi identity in usual GR, is obtained by 
applying the operator of covariant derivative with respect to $A$ to
the equation (\ref{eqs-f}). In view of $\cD F^i=0$ identically, we get
\be\label{bianchi}
\cD (\Lambda \delta^{ij} + \Psi^{ij} + \phi \delta^{ij}) B^j = 8\pi G\, \cD T^i,
\ee
In usual GR, contracting this
equation in a certain way gets rid of the left-hand-side, and gives the
equation of energy conservation. We would like to obtain an analog of
this equation for theory (\ref{action}).

\bigskip
\noindent{\bf Computation.} 
To compute the covariant derivative of the quantity on the left-hand-side
of (\ref{bianchi}) we need to know how $\cD$ acts on 
the basic rank 2 internal spinors $\tilde{X}_\pm,\tilde{X}$ as
well as on the 2-forms $B_\pm,B$. The action of 
$\cD$ on tilded ``internal'' rank 2 spinors is given by:
\be
\cD \tilde{X}_- = -\tilde{A} \tilde{X}_- - 2\tilde{A}_+ \tilde{X},
\qquad 
\cD \tilde{X}_+ = \tilde{A}\tilde{X}_+  + 2\tilde{A}_- \tilde{X}, 
\qquad
\cD \tilde{X}  = \tilde{A}_- \tilde{X}_- - \tilde{A}_+ \tilde{X}_+,
\ee
which is easy to show using the formulae (\ref{ta}) for covariant
derivatives of the basic spinors. 

The action of $\cD$ on 2-forms $B_\pm,B$ is that of the derivative
operator $\nabla$ compatible with the metric defined by $l,n,m,\mb$.
The most economic way to compute this action is using spinor
representation. In this representation the self-dual 2-forms 
$B_\pm,B$ become the matrices (\ref{b1}). The action of the 
operator $\nabla$ on them can be obtained from relations 
(\ref{c-ab}) and (\ref{nB}). We have the following formulae:
\be
(\nabla B_-) = B_- \tilde{A} - B\tilde{A}_-, \qquad
(\nabla B_+) = B \tilde{A}_+ - B_+\tilde{A}, \qquad
(\nabla B) = 2(B_- \tilde{A}_+ - B_+ \tilde{A}_-),
\ee
where matrix multiplication on the right-hand-side is implied. Using these
formulae, and passing into the spinor representation by replacing
3-forms of the type $A\wedge B$ by the spinor quantity $BA$ 
(matrix multiplication implied) we get the following 3 components
of the ``Bianchi'' identities:
\be\label{id-1}
B_+ \nabla\alpha + B_- \nabla(\beta+\phi) + 
\alpha(B\tilde{A}_+ - 2B_+\tilde{A})-3\beta B\tilde{A}_- = 8\pi G (\cD T)_-,
\\ \nonumber
B_- \nabla\alpha + B_+ \nabla(\beta+\phi) + 
\alpha(2B_-\tilde{A}-B\tilde{A}_-)+3\beta B\tilde{A}_+ = 8\pi G (\cD T)_+,
\\ \nonumber
B\nabla(\phi-2\beta) + 2\alpha(B_-\tilde{A}_- - B_+\tilde{A}_+)
-6\beta (B_-\tilde{A}_+ - B_+\tilde{A}_-) = 8\pi G (\cD T),
\ee
where on the right hand side of these equations one finds
the projections of the quantity $(\cD T)^{AB}$ onto the 3 rank 2 spinors 
$\tilde{X}_\pm,\tilde{X}$.

\bigskip
\noindent{\bf Contracted ``Bianchi'' identity.}
Let us now multiply the first of the equations in (\ref{id-1})
by $B_+$, the second of them by $B_-$ and add the results. We get:
\be\label{id-2}
(B_+ B_+ + B_- B_-) \nabla\alpha + (B_+ B_- + B_- B_+)\nabla(\beta+\phi)
+ \alpha(B_+ B\tilde{A}_+ - B_- B\tilde{A}_-) - 3\beta 
(B_+ B\tilde{A}_- - B_- B\tilde{A}_+) = \\ \nonumber
8\pi G (B_+  (\cD T)_- + B_-  (\cD T)_+).
\ee
Let us multiply the third equation in (\ref{id-1}) by $B/2$. We get:
\be
(1/2) BB \nabla(\phi-2\beta)+ \alpha(BB_-\tilde{A}_- - BB_+\tilde{A}_+)
-3\beta (BB_-\tilde{A}_+ - BB_+\tilde{A}_-) = 8\pi G ((1/2)B(\cD T)).
\ee
Now, subtracting this equation from (\ref{id-2}), and using the
fact that $BB_+ + B_+B=0, BB_- + B_- B=0$ we get:
\be\label{id-3}
(B_+ B_+ + B_- B_-) \nabla\alpha + (B_+ B_- + B_- B_+)\nabla(\beta+\phi)
- (1/2) BB \nabla(\phi-2\beta) = \\ \nonumber
8\pi G (B_+  (\cD T)_- + B_-  (\cD T)_+ - (1/2)B(\cD T)).
\ee
It is now elementary to check that the quantity on the left-hand-side
of this equation is zero. Indeed, using $B_+ B_+ + B_- B_- = -2a, 
B_+ B_- + B_- B_+ = -(1+a^2), BB = c^2$ we get for the left-hand-side
of (\ref{id-3}):
\be
-2a \nabla\alpha - (1+a^2)\nabla(\beta+\phi) - (c^2/2) \nabla(\phi-2\beta) = 
-2a \nabla\alpha - (1+a^2 - c^2)\nabla\beta - (1+a^2-c^2/2)\nabla\phi.
\ee 
However, this obviously vanishes in view of (\ref{metricity-rels}).
Thus, we have established the following 
of ``energy-momentum'' conservation law for theory (\ref{action}):
\be\label{energy}
B_+  (\cD T)_- + B_-  (\cD T)_+ - (1/2)B(\cD T) = 0.
\ee
Using the fact that $\cD B=0$, we can rewrite this identity as:
\be
\nabla B T = 0,
\ee
or, re-introducing all indices:
\be
\nabla^{BB'} B_{AB}^{CD} T_{CD\,A'B'} = 0.
\ee
Thus, we see that the energy conservation law in this theory takes the usual form, with the
stress-energy tensor that is conserved being $T_{ab}=(BT)_{ab}$. The 2-form $B$ appears in
the expression for the quantity $T_{ab}$ as it is necessary to convert the pair $AB$ of
``internal'' indices of $T^{AB}_{A'B'}$ into a pair of spacetime spinor indices. After this
is done one gets a quantity with only spacetime indices, and it is this quantity that is
conserved in the usual way.

\section*{Acknowledgements}

It is a pleasure to thank Laurent Freidel, Andrei Mikhailov, Yuri Shtanov and Sergey Winitzki for stimulating 
discussions. The author was partially supported by an EPSRC advanced fellowship and
by Perimeter Institute for Theoretical Physics.

\section*{Appendix: The ``metricity'' equations}

The main novelty of the theory under consideration as compared to Pleba\'nski one is
in the content of ``metricity'' equations that follow by varying the action
with respect to the Lagrange multiplier field $\Psi$. As we have already discussed
in the main text, these equations should be interpreted as determining the
components of $\Psi$ in terms of the components of the $B$ field. For
practical considerations, however, it is more convenient to interpret
these equations as determining the 5 components of the $B$ field (that
$B$ describes in addition to a metric) in terms of the $\Psi$ field. For
this interpretation one needs to introduce a metric. The most natural
way to do this, also suggested by what happens in the case of Pleba\'nski
gravity theory, is to declare the triple of two-forms $B_\pm, B$ to
be the self-dual two-forms with respect to some metric, which
then determines the metric up to a conformal factor, in the way
discussed in the main text. 

In this appendix we prove that using the ${\rm SL}(2)$ freedom in
choosing the basis $l,n,m,\mb$, any solution of the
equations (\ref{metricity-1}) can be brought into the form (\ref{B*}).

As a preliminary, let us consider the question of how a general self-dual 2-form
\be
C = c_1 m\wedge l + c_2 n\wedge \mb + c_3 (l\wedge n - m\wedge \mb)
\ee
can be simplified by rotating the basis $l,n,m,\mb$. Thus, let us consider
transformations that act on $l,m$ and involve the other one-forms $n,\mb$:
\be\label{app-inhom}
l\to l+\alpha n +\beta \mb, \qquad m\to m+ \gamma n +\delta \mb.
\ee
Simple analysis shows that there is a way to perform this
transformation that no anti-self-dual part is introduced.
This requires: $\alpha=\delta, \beta=\gamma=0$. Under such
a transformation the 2-form $C$ changes to
\be\label{app-4}
C\to c_1 m\wedge l + (c_1 \alpha^2 + c_2 - 2\alpha c_3) 
n\wedge \mb + (c_3-c_1 \alpha) (l\wedge n- m\wedge\mb).
\ee
Thus, choosing $\alpha=c_3/c_1$ we can eliminate the last term and (using
an additional rescaling of the one-forms $l,m$) 
bring a general self-dual 2-form into the form:
\be\label{app-2}
C= m\wedge l + c n\wedge \mb,
\ee
We emphasise that in order to achieve this form only the transformations of
the one-forms $l,m$ were used. One can show by a similar argument
that any 2-form $C'$ can be brought into the form: 
$C'=n\wedge\mb + c' m\wedge l$ by acting on the one-forms $n,\mb$ only.

Let us now consider three self-dual
2-forms $B_\pm,B$ satisfying (\ref{metricity-1}). We can 
use the available freedom in choosing the basis $l,n,m,\mb$ to
bring these 3-forms into the desired form (\ref{B*}). To achieve
this we first bring the self-dual 2-forms $B_\pm$ into their ``canonical''
forms (as discussed above) by acting on the pairs $l,m$ via 
$l\to l + \alpha n, m\to m+\alpha \mb$ and on $n,\mb$
via $n\to n+\alpha ' l, \mb\to \mb+\alpha' m$ 
correspondingly, as well as rescaling $l,n,m,\mb$. This way we achieve
\be\label{app-3}
B_+ = m\wedge l + a\,n\wedge \mb, \qquad B_- = n\wedge \mb + a'\, m\wedge l.
\ee 
Importantly, under these transformations the self-dual 2-form $B$ remains
self-dual, as is clear from (\ref{app-4}). Given the 2-forms (\ref{app-3})
together with a general self-dual 2-form $B$ it is easy to see
that the metricity equations of the second line of 
(\ref{metricity-1}) imply that $B$ is of the form
\be
B=c(l\wedge n- m\wedge\mb).
\ee
The two equations of the
first line of (\ref{metricity-1}) imply, in particular, that $a=a'$.
This finishes our proof of the validity of the 
representation (\ref{B*}) for $B$. 

\section*{Appendix: Connection $A$ without spinors}

Sometimes it is more convenient to avoid spinors, and do computations
directly with forms. Here we give the expressions for the spin-coefficients
that are obtained this way. Thus, using the expression (\ref{B*}) for $B$ 
and decomposing $A$ into the basis of one-forms, we can easily compute 
the projections of $C_\pm,C$ onto the basis one-forms. We have:
\be
(C_-\wedge l) = c A_{-n} - aA_m, \qquad 
(C_-\wedge n) = - c A_{-l} + A_{\mb}, \\ \nonumber
(C_-\wedge m) = - c A_{-\mb} + aA_l, \qquad
(C_-\wedge \mb) = c A_{-m} - A_n.
\ee
\be
(C_+\wedge l) = A_m - c A_{+n}, \qquad 
(C_+\wedge n) = -a A_{\mb} + cA_{+l}, \\ \nonumber
(C_+\wedge m) = -A_l + cA_{+\mb}, \qquad
(C_+\wedge \mb) = aA_n-cA_{+m}.
\ee
\be
(C\wedge l) = A_{-m} - a A_{+m}, \qquad 
(C\wedge n) = A_{+\mb} - aA_{-\mb}, \\ \nonumber
(C\wedge m) = -A_{-l} + aA_{+l}, \qquad
(C\wedge \mb) = -A_{+n}+aA_{-n}.
\ee

Assuming $C_\pm,C$ are given, it is easy to write down the solution 
for $A$. We get:
\be
A_{-l}=-\frac{1}{1-a^2}
\left( (C\wedge m) - \frac{a}{c}((C_++aC_-)\wedge n)\right), \quad
A_{-n}=-\frac{1}{1-a^2}
\left( a (C\wedge \mb) - \frac{1}{c}((C_- + aC_+)\wedge l)\right), \\
\nonumber
A_{-m}=\frac{1}{1-a^2}
\left( (C\wedge l) - \frac{a}{c}((C_+ +aC_-)\wedge\mb)\right), \quad
A_{-\mb}=\frac{1}{1-a^2}
\left( a (C\wedge n) - \frac{1}{c}((C_- + aC_+)\wedge m)\right).
\ee
\be
A_{+l}=-\frac{1}{1-a^2}
\left( a (C\wedge m) - \frac{1}{c}((C_+ +aC_-)\wedge n)\right), \quad
A_{+n}=-\frac{1}{1-a^2}
\left( (C\wedge \mb) - \frac{a}{c}((C_- + aC_+)\wedge l)\right), \\
\nonumber
A_{+m}=\frac{1}{1-a^2}
\left( a (C\wedge l) - \frac{1}{c}((C_+ +aC_-)\wedge\mb)\right), \quad
A_{+\mb}=\frac{1}{1-a^2}
\left( (C\wedge n) - \frac{a}{c}((C_- + aC_+)\wedge m)\right).
\ee
\be
A_{l}=\frac{1}{1-a^2}
\left( c (C\wedge n) - ((C_+ +aC_-)\wedge m)\right), \quad
A_{n}=\frac{1}{1-a^2}
\left( c (C\wedge l) - ((C_- + aC_+)\wedge \mb)\right), \\
\nonumber
A_{m}=-\frac{1}{1-a^2}
\left( c (C\wedge \mb) - ((C_+ +aC_-)\wedge l)\right), \quad
A_{\mb}=-\frac{1}{1-a^2}
\left( c (C\wedge m) - ((C_- + aC_+)\wedge n)\right).
\ee
These relations should be compared to (\ref{a-pm}), (\ref{a}).

To obtain the spin coefficients $\tilde{A}$ we can use 
the following identities:
\be
-(\tilde{C}_+ +a\tilde{C}_-) = (1+a^2) d(n\wedge \mb) + 2a d(m\wedge l) + da\wedge B_-, 
\\ \nonumber
-(\tilde{C}_- +a\tilde{C}_+) = (1+a^2) d(m\wedge l) + 2a d(n\wedge \mb) + da\wedge B_+.
\ee
This gives:
\be
-((\tilde{C}_+ +a\tilde{C}_-)\wedge l) = (1+a^2)( (d\mb)_{m\mb}- (dn)_{nm}) 
-2a (dl)_{n\mb} + a (da)_m, \\ \nonumber
-((\tilde{C}_+ +a\tilde{C}_-)\wedge n) = - (1+a^2) (dn)_{ml} + 2a( (dm)_{m\mb} + (dl)_{l\mb})
- (da)_{\mb}, \\ \nonumber
-((\tilde{C}_+ +a\tilde{C}_-)\wedge m) = (1+a^2)( - (d\mb)_{l\mb} - (dn)_{ln}) 
-2a (dm)_{n\mb} - a (da)_l, \\ \nonumber
-((\tilde{C}_+ +a\tilde{C}_-)\wedge \mb) = - (1+a^2) (d\mb)_{ml} + 2a( (dm)_{nm} - (dl)_{ln})
+ (da)_{n}.
\ee
\be
-((\tilde{C}_- +a\tilde{C}_+)\wedge l) = - (1+a^2) (dl)_{n\mb}
+2a (-(dn)_{nm} +(d\mb)_{m\mb}) + (da)_m, \\ \nonumber
-((\tilde{C}_- +a\tilde{C}_+)\wedge n) = (1+a^2)( (dm)_{m\mb} - (dl)_{n\mb}) 
-2a (dn)_{ml} - a (da)_{\mb}, \\ \nonumber
-((\tilde{C}_- +a\tilde{C}_+)\wedge m) = - (1+a^2) (dm)_{n\mb}
+2a (-(dn)_{ln} -(d\mb)_{l\mb}) - (da)_l, \\ \nonumber
-((\tilde{C}_- +a\tilde{C}_+)\wedge \mb) = (1+a^2)( (dm)_{nm} - (dl)_{ln}) 
-2a (d\mb)_{ml} + a (da)_{n}.
\ee
The notation that we used here is almost self-explanatory. Thus, e.g.
$dm = (dm)_{nm} n\wedge m + \ldots$, where the dots stand for the other
terms. Finally, we get for the projections of the 3-form $\tilde{C}$:
\be
- 2(\tilde{C}\wedge l) = - c(dl)_{m\mb} + c(dm)_{nm} + c(d\mb)_{n\mb} + (dc)_n,
\\ \nonumber
- 2(\tilde{C}\wedge n) = - c(dn)_{m\mb} + c(dm)_{ml} - c(d\mb)_{l\mb} - (dc)_l,
\\ \nonumber
- 2(\tilde{C}\wedge m) =  c(dl)_{l\mb} + c(dn)_{n\mb} + c(dm)_{ln} - (dc)_{\mb},
\\ \nonumber
- 2(\tilde{C}\wedge \mb) =  c(dl)_{ml} - c(dn)_{nm} + c(d\mb)_{ln} + (dc)_m.
\ee

\end{document}